\documentclass[twocolumn, trackchanges]{aastex63}
\usepackage[utf8]{inputenc}

\usepackage{siunitx}
\usepackage{natbib}
\usepackage{graphicx}
\usepackage{xcolor}
\usepackage{hyperref}
\begin{document}
\title{Consistent analysis of the AGN LF in X-ray and MIR in the XMM-LSS field}

\author[0000-0001-9921-7100]{Jack Runburg}
\affiliation{\mbox{Department of Physics \& Astronomy,
University of Hawai`i, Honolulu, HI 96822, USA}}

\author[0000-0003-1748-2010]{Duncan Farrah}
\affiliation{\mbox{Department of Physics \& Astronomy,
University of Hawai`i, Honolulu, HI 96822, USA}}
\affiliation{\mbox{Institute for Astronomy, 2680 Woodlawn Drive, University of Hawaii, Honolulu, HI 96822, USA}}

\author[0000-0002-1917-1200]{Anna Sajina}
\affiliation{\mbox{Department of Physics \& Astronomy, Tufts University, Medford, MA 02155, USA}}

\author[0000-0002-3032-1783]{Mark Lacy}
\affiliation{National Radio Astronomy Observatory, 520 Edgemont Road, Charlottesville, VA 22903, USA}

\author[0000-0002-5677-318X]{Jenna Lidua}
\affiliation{\mbox{Department of Physics \& Astronomy,
University of Hawai`i, Honolulu, HI 96822, USA}}

\author[0000-0003-0917-9636]{Evanthia Hatziminaoglou}
\affiliation{\mbox{European Southern Observatory, Karl-Schwarzschild-Str. 2, 85748 Garching bei M\"unchen, Germany}}

\author[0000-0002-0167-2453]{W.N. Brandt}
\affiliation{\mbox{Department of Astronomy and Astrophysics, 525 Davey Lab, The Pennsylvania State University, University Park, PA 16802, USA}}
\affiliation{\mbox{Institute for Gravitation and the Cosmos, The Pennsylvania State University, University Park, PA 16802, USA}}
\affiliation{\mbox{Department of Physics, 104 Davey Laboratory, The Pennsylvania State University, University Park, PA 16802, USA}}

\author[0000-0002-4945-5079]{Chien-Ting J. Chen}
\affiliation{\mbox{Astrophysics Office, NASA Marshall Space Flight Center, ZP12, Huntsville, AL 35812, USA}}

\author[0000-0003-1991-370X]{Kristina Nyland}
\affiliation{National Research Council, resident at the Naval Research Laboratory, Washington, DC 20375, USA}

\author[0000-0002-1114-0135]{Raphael Shirley}
\affiliation{\mbox{Astronomy Centre, Department of Physics \& Astronomy, University of Southampton, Southampton, SO17 1BJ, UK}}
\affiliation{\mbox{Institute of Astronomy, University of Cambridge, Madingley Road, Cambridge, CB3 0HA, UK}}

\author[0000-0002-9548-5033]{D.L. Clements}
\affiliation{\mbox{Physics Department, Imperial College London, Blackett Laboratory, Prince Consort Road, London, SW7 2AZ, UK}}

\author{Lura K. Pitchford}
\affiliation{\mbox{Department of Physics, Virginia Tech, Blacksburg, VA 24061, USA}}

\date{\today}

\begin{abstract}
The luminosity function (LF) of active galactic nuclei (AGN) probes the history of supermassive black hole assembly and growth across cosmic time. 
To mitigate selection biases, we present a consistent analysis of the AGN LFs derived for both X-ray and mid-infrared (MIR) selected AGN in the XMM-Large Scale Structure (XMM-LSS) field.
There are 4268 AGN used to construct the MIR luminosity function (IRLF) and 3427 AGN used to construct the X-ray luminosity function (XLF), providing the largest census of the AGN population out to $z=4$ in both bands with significant reduction in uncertainties. 
We are able for the first time to see the knee of the IRLF at $z>2$ and observe a flattening of the faint-end slope as redshift increases.
The bolometric luminosity density, a proxy for the cosmic black hole accretion history, computed from our LFs shows a peak at $z\approx2.25$ consistent with recent estimates of the peak in the star formation rate density (SFRD). 
However, at earlier epochs, the AGN luminosity density is flatter than the SFRD.
If confirmed, this result suggests that the build up of black hole mass outpaces the growth of stellar mass in high mass systems at $z\gtrsim 2.5$.
This is consistent with observations of redshift $z\sim 6$ quasars which lie above the local $M-\sigma$ relationship.
The luminosity density derived from the IRLF is higher than that from the XLF at all redshifts. 
This is consistent with the dominant role of obscured AGN activity in the cosmic growth of supermassive black holes.
\end{abstract}

\keywords{Active galactic nuclei, luminosity functions, galaxy formation}

\section{Introduction}
\label{sec:intro}
Active galactic nuclei (AGN) are the central regions of active galaxies that produce a huge amount of radiation from radio to $\gamma$-rays.
This radiation originates from accretion onto supermassive black holes. 
Different spectra are observed from AGN due to, e.g., accretion histories, obscuring material, and orientation of the galaxy along the line of sight.
Because this accretion is tied to the central black hole, the luminosity of AGN can provide insight into galaxy structure and dynamics as well as the cosmological evolution of galaxies and their formation.
The comoving AGN luminosity density peaks around $z\sim 2$ \citep[e.g.,][]{aird_evolution_2010} as does the star formation rate \citep[e.g.,][]{madau_cosmic_2015}, suggesting that a significant fraction of galaxy formation and assembly happens at high redshifts ($1 < z < 6$) due to high rates of star formation and black hole mass accretion.

Luminosity functions are a useful tool to study the formation and evolution of galaxies.
Complete surveys with well-understood selection biases are needed to construct accurate AGN LFs; care must be taken, however, in quantifying biases and completeness.
Optical and X-ray surveys have been used in abundance to constrain the AGN LF.
However, these surveys are less successful at identifying heavily obscured AGN \citep{hickox_obscured_2018}.
A complete census of AGN--including obscured AGN--is necessary to accurately determine the AGN LF and test models of galaxy formation and evolution \citep{shen_bolometric_2020}.
Within the AGN unification scheme, obscuration is due to intervening material between the observer and the central engine.
High-energy radiation that would otherwise indicate the presence of an AGN, quasar or blazar is reprocessed as it travels through the coaxial dusty torus and then is reemitted in the infrared.
Without corrections for obscuration, any X-ray AGN sample will be necessarily incomplete, representing the unobscured AGN population and excluding the obscured population, whereas a mid-IR AGN sample represents a mixture of obscured and unobscured AGN that dominate over the dust emission of the host galaxy.
Therefore, observations in the infrared provide a unique window into the population of obscured AGN.

Infrared surveys are arguably ideal for obtaining a complete census of the AGN population, but are more prone to contamination by non-AGN objects for any choice of selection criteria.
Previous work  has used first \textit{Spitzer} AGN \citep{lacy_spitzer_2015},  and later a combination or \textit{Spitzer} and \textit{WISE} AGN \citep{2018ApJ...861...37G} to constrain the IRLF.
However, deeper and larger surveys are needed to constrain the bright end of the IRLF out to higher redshifts.
Additionally, selecting AGN in IR bands is complicated and redshift-dependent.
Other astrophysical sources (e.g. star forming galaxies) can have similar or larger emission in the IR,  resulting in trade-offs in reliability and completeness in AGN selection techniques \citep[e.g.,][]{2014A&A...562A.144M,2020NatAs...4..352L,hatziminaoglou_sloan_2005}
Thus, any IR-selected AGN sample is likely to be either incomplete or contaminated by non-AGN objects to some degree.
Multi-band observations in the same survey area can be used to better select AGN and provide better estimates for photometric redshifts \citep{laigle2016ApJS..224...24L, duncanI2018MNRAS.473.2655D, duncanII2018MNRAS.477.5177D, salvato_many_2019}.
Since our sample includes $\sim 5000$ AGN at $z\gtrsim 1$ over a survey area of \SI{11}{deg^2}, our sample necessarily includes photometric redshifts for many of the selected AGN.\footnote{Obtaining spectroscopic redshifts for obscured (i.e. optically-faint) AGN is hindered by the relationship between optical line strength and AGN luminosity (see, e.g., Fig.~12 of \cite{lacy_spitzer_2013}).}

In contrast to IR-selection of AGN, selection of AGN in the X-ray is quite efficient \citep[e.g.,][]{brandt2015A&ARv..23....1B}.
With few exceptions, galaxies produce X-ray luminosities less than (i.e. $L_X < \SI{1e42}{erg~s^{-1}}$) that characteristic of AGN \citep{mineo_x-ray_2012}.
However, some Compton-thick AGN (i.e. AGN with column densities $N_H$ exceeding \SI{1.5e24}{cm^{-2}}) will not be efficiently selected in the X-ray and so the X-ray selected AGN sample is bound to be incomplete \citep{hickox_obscured_2018}.
Construction of the LF in both X-ray and mid-IR bands provides a powerful crosscheck that sample incompleteness has been adequately addressed in both bands.

In this paper, we present a consistent analysis of the IRLF and XLF in the XMM-LSS field, including multiple corrections for incompleteness and new methodology for addressing photometric redshift uncertainties.
In \S\ref{sec:sample}, we describe our data and AGN selection criteria.
\S\ref{sec:methods} describes the construction of the binned luminosity function using the $1/V_\mathrm{max}$ method as well as our incompleteness corrections and the probability distribution functions (PDFs) we use to account for photometric redshift uncertainties.
Our results are presented in \S\ref{sec:results} and discussed in \S\ref{sec:discussion}.
For all relevant calculations, we assume a flat cosmology with $H_0=\SI{67.4}{km~s^{-1}~Mpc^{-1}}$, $\Omega_m = 0.32$ and $\Omega_\Lambda=0.68$ \citep{planck_collaboration_planck}.

\begin{figure*}[tphb]
\centering
\includegraphics[width=0.9\textwidth]{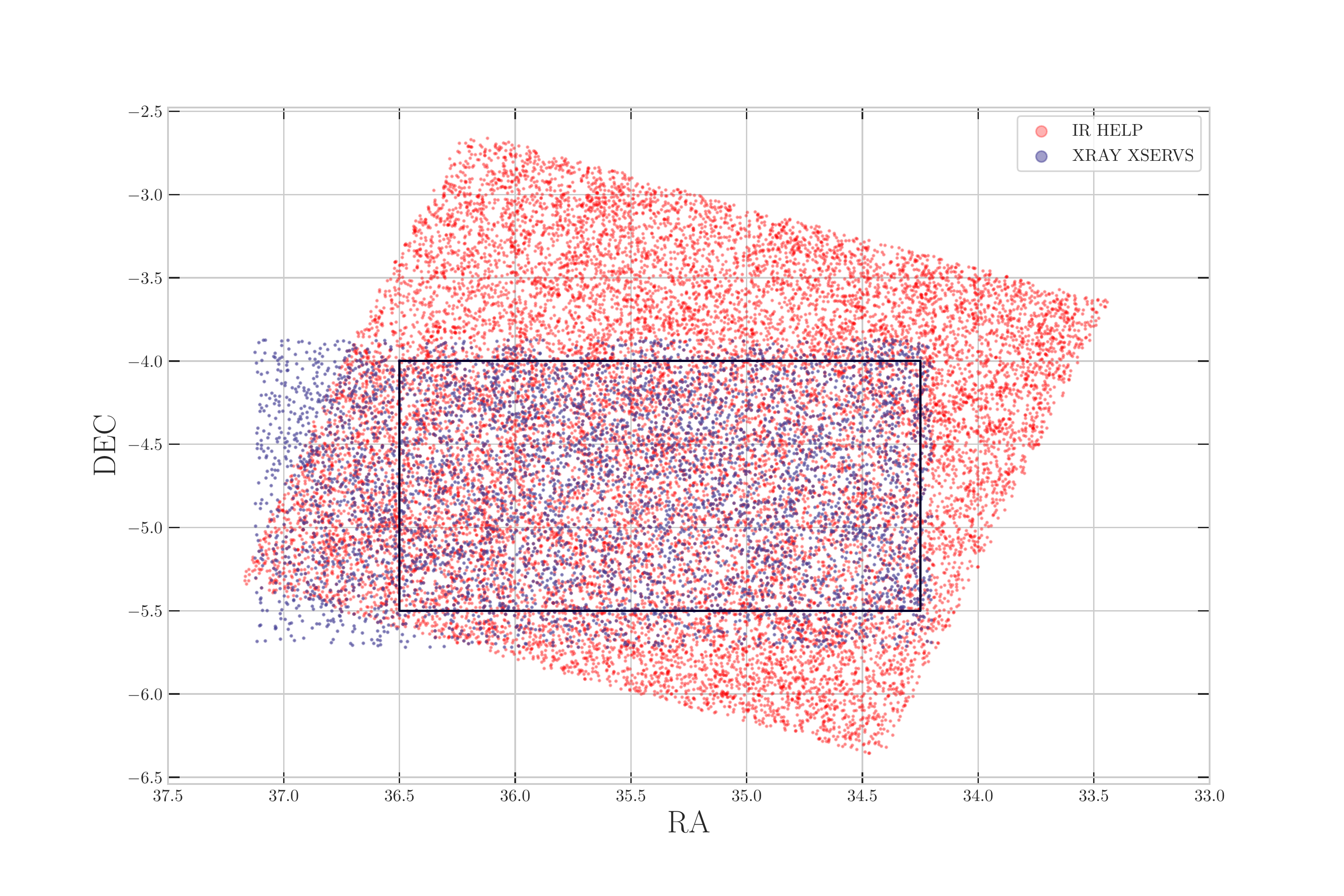}
\caption{Footprint of the HELP catalog (red) and XMM-SERVS catalog (blue) in the XMM-LSS field. There are 30,271 IR sources in the HELP catalog and 5242 X-ray sources in the XMM-SERVS catalog. The black box indicates the subregion used for the comparative sample (described in \S\ref{sec:crossmatch}).}
\label{fig:xmatch_sample}
\end{figure*}

\section{Data}
\label{sec:sample}
The XMM-LSS field is a $\sim\SI{11}{deg^2}$ field in the near-equatorial southern Galactic hemisphere.
Multi-wavelength coverage from radio to X-ray makes this field an ideal candidate for comparing the LF constructed in different bands.
In particular, we use the \textit{Herschel} Extragalactic Legacy Project (HELP) catalog \citep{shirley_help_2019, shirley_help_2021} to construct the IRLF and the XMM-Spitzer Extragalactic Representative Volume Survey (XMM-SERVS) catalog \citep{chen_xmm-servs_2018} to construct the XLF.
Fig.~\ref{fig:xmatch_sample} shows the sources from both catalogs as well as a common region in which we construct the LF on a comparative sample from both catalogs.

In this section, we detail the X-ray and IR samples we use to construct our LFs. 
Additionally, we discuss the selection criteria, redshifts, and luminosities used in our analysis.

\subsection{IR data}
We use MIR fluxes from the HELP \citep{shirley_help_2021} XMM-LSS catalog covering $\sim \SI{9.0}{deg^2}$ of the XMM-LSS field.
HELP standardizes and combines data in optical, near-infrared, and mid-infrared bands across 23 different survey fields. 
In particular we use IRAC1 and IRAC2 photometry from SERVS \citep{mauduit_spitzer_2012}, IRAC3 and IRAC4 photometry from SWIRE \citep{lonsdale_swire_2003}, and \SI{24}{\micro m} photometry from SWIRE.
We restrict our sample to the sources in the XMM-LSS field with photometry in all four IRAC bands and MIPS 24 micron photometry, as well as a redshift yielding a sample of 18,328 sources.
Since the survey is flux-limited ($5\sigma$ sensitivity level is \SI{27.5}{\micro Jy}) by the \SI{5.8}{\micro\meter} (IRAC3) band, our IRLF is constructed using the \SI{5.8}{\micro m} fluxes.

\subsection{X-ray data}
The XMM-LSS field has been targeted by multiple X-ray survey campaigns such as the original XMM-LSS survey, XMDS, UDS, XMM-XXL, and XMM-SERVS.
For this work we use the XMM-SERVS catalog \citep{chen_xmm-servs_2018} covering the center $\sim \SI{5.3}{deg^2}$ of the XMM-LSS field with a homogeneous $\sim \SI{50}{ks}$ XMM-Newton observations combining
new and archival data in the field.
The XMM-SERVS catalog \citep{chen_xmm-servs_2018} consists of 5242 X-ray point sources, combining many \textit{XMM-Newton} observations from several surveys (XMM-LSS, XMDS, XMM-XXL, archival, and substantial new data) with flux limits of \SI{1.7e-15}{erg~cm^{-2}~s^{-1}} for the soft band (\SIrange[range-phrase=--, range-units=single]{0.5}{2}{keV}), \SI{1.3e-14}{erg~cm^{-2}~s^{-1}} for the hard band (\SIrange[range-phrase=--, range-units=single]{2}{10}{keV}), and \SI{6.5e-15}{erg~cm^{-2}~s^{-1}} for the full band (\SIrange[range-phrase=--, range-units=single]{0.5}{10}{keV}) over 90\% of the \SI{5.3}{deg^2} area.
The catalog also includes crossmatches in the optical, near-IR, and mid-IR.
We restrict our sample to the sources with photometry in soft, hard and full bands and that have a redshift.
Our XLF is constructed using the full band fluxes.

\subsection{Redshifts}
Both XMM-SERVS and HELP provide a substantial increase  (greater than a factor of $\sim 2$) in the sample size of AGN over previous studies. 
However, the need for accurate redshift estimates is just as important to compute a reliable LF.
Spectroscopic redshifts are preferred but often severely limit the sample size.
Photometric redshifts are typically obtained by fitting templates to the observed multi-wavelength spectral energy distribution (SED) of a source.
Depending on the wavelength and redshift, components of the AGN SED may resemble those of other astrophysical objects.
Similarly, untangling the AGN contribution to the total emission can be challenging because the host galaxy may contribute a large fraction of the total emission in certain wavelength ranges \citep{salvato_many_2019}.
Our data have a mix of photometric and spectroscopic redshifts.
To quantify the reliability of our photometric redshifts, we define $\sigma_{\rm nmad}\equiv1.48\times\mathrm{median}(\Delta z/(1 + z_{\rm spec}))$ and outlier fraction, $f_{\rm out}$, where outliers satisfy $\Delta z > 3\sigma_{\rm nmad}$.
Variability of AGN \citep{kozlowski2016ApJ...826..118K} is also a problem for photo-z measurements, especially when observations are taken in different bands at different epochs.
This variability may contribute to scatter of the redshifts but most of the optical emission comes from the host galaxy and thus should be relatively insensitive to changes in the AGN emission.
Additionally, given the deep spectroscopic coverage of this survey region, the variability effects are more likely to be found in type 1 AGN that already have spectroscopic redshifts.

The HELP catalog includes spectroscopic redshifts when available.
Photometric redshifts are obtained using standard SED template fitting \citep{duncanI2018MNRAS.473.2655D} with the default \texttt{EAZY} \citep{brammer_eazy_2008} templates, stellar-only templates, and many templates of different galaxy spectral types with contributions from AGN and QSOs.
These estimates are improved by using a classifier trained on spectroscopically-identified AGN that accounts for the priors and is calibrated to the uncertainties for the known AGN \citep{duncanII2018MNRAS.477.5177D}; the two estimates are combined to form a consensus estimate.
% With these combined estimates, the normalized median absolute deviation is $\sigma_\mathrm{NMAD}\sim0.1$ with an outlier fraction of $\sim25\%$.
For IR selected AGN (as described in \S\ref{sec:selectir}), the normalized median absolute deviation is $\sigma_\mathrm{NMAD}=0.08$ with an outlier fraction of $\sim24\%$.
Photometric redshifts are available for $5847/5918\approx 99\%$ of the AGN.
The roughly 80 AGN without photometric redshift all have spectroscopic redshifts.
Spectroscopic redshifts are available for $2283/5918\approx 39\%$ of AGN.

The XMM-SERVS catalog similarly includes spectroscopic redshifts from various spectroscopic surveys.
Spectroscopic redshifts are present for $1767/5242\approx 34\%$ of the sources.
High quality photometric redshifts are available for a \SI{4.5}{\deg^2} region in the field based on the forced-photometry catalog from \cite{pforr_photometric_2019}.
The SED-fitting code \texttt{EAZY} is used with the default galaxy templates and an additional obscured AGN template.
The normalized median absolute deviation is $\sigma_\mathrm{NMAD}=0.05$ with outlier fraction $f_\mathrm{outlier}=20.0\%$.
Photometric redshifts are available for roughly $3419/5242\sim40\%$ of the sources.

\begin{figure}[htpb]
\centering
\includegraphics[width=0.95\columnwidth]{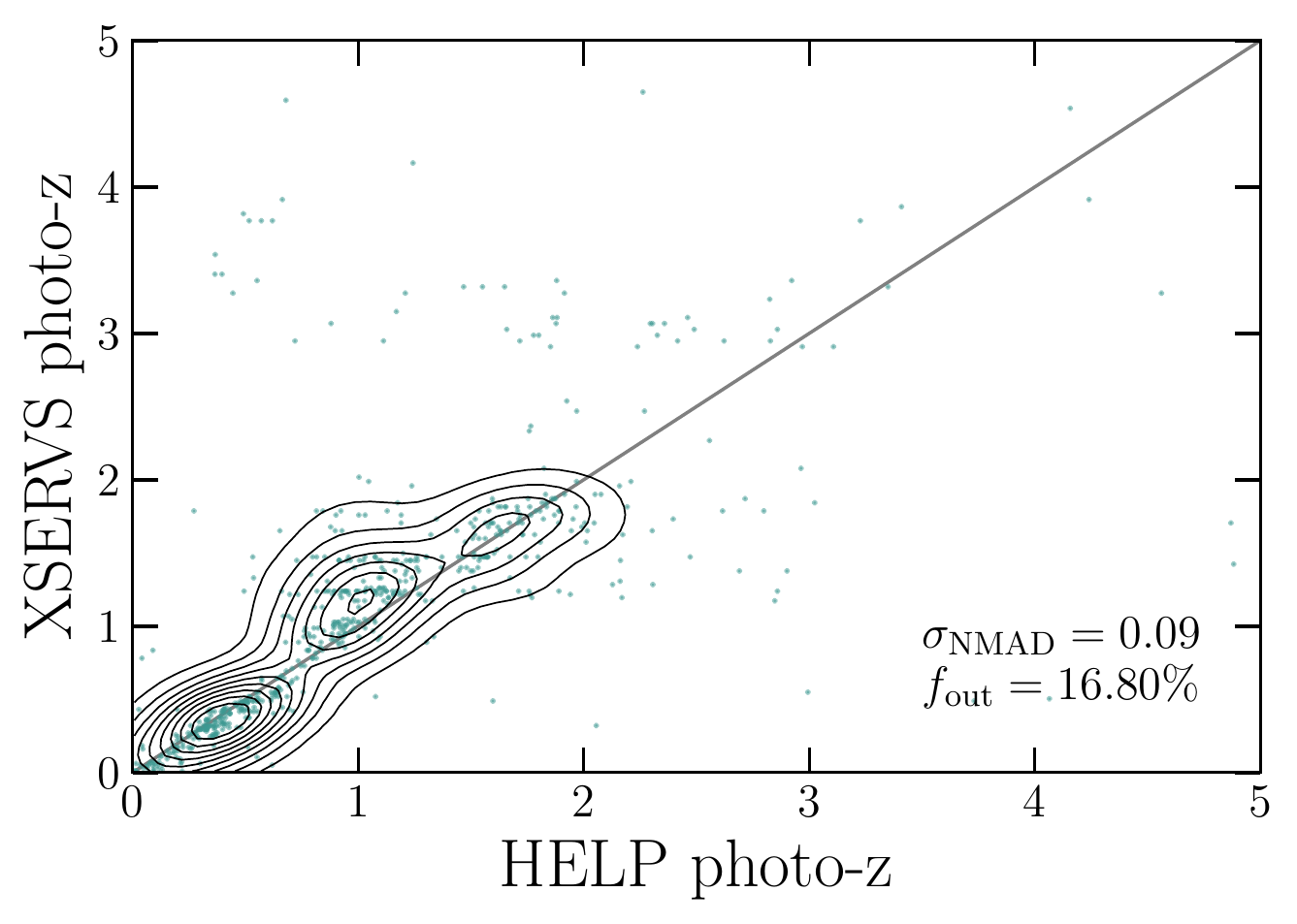}
\caption{Comparison of photometric redshifts for sources in the comparative sample with photo-zs from XMM-SERVS and HELP. There are 811 sources. Contours are plotted for the distribution using a KDE with bandwidth corresponding to the normalized median absolute deviation $\sigma_\mathrm{NMAD}=0.09$. The contours trace out the regions of equal likelihood starting from the outside at 10\% and increasing in 10\% increments up to 90\% in the very center.}
\label{fig:photozcomp}
\end{figure}

We recognize that using two sets of photometric redshifts adds uncertainty.
Our analysis includes checks to understand the effects of the different photometric redshifts.
The value of $\sigma_\mathrm{NMAD}$ for each catalog are comparable suggesting that neither set of photometric redshifts is significantly better than the other.
In Fig.~\ref{fig:photozcomp}, we plot a comparison of the photometric redshifts from HELP and XMM-SERVS for the objects in the comparative region (described further in \S\ref{sec:crossmatch}).
To get an idea of the difference between the two photometric redshifts, we include likelihood contours that indicate how likely a source is to fall within a given region--the outermost contour encloses 90\% of the sources (i.e. 10\% of the sources are outside of the outermost contour) and the innermost contour encloses 10\% of the sources.
Many of the sources have comparable redshifts from HELP and XMM-SERVS, but a significant fraction are dissimilar, stemming from a failure of at least one of the photometric redshift fitting routines.
For sources without spectroscopic redshifts, we have no systematic way of identifying which photometric estimate is more accurate (and for sources that do not have a crossmatch in the other catalog, we have only one photometric redshift estimate to rely on).
We attempt to address these uncertainties in \S\ref{sec:mc}.
Ultimately, we want to ensure that using different sets of redshifts does not bias our estimate of the LF; we address this issue in \S\ref{sec:results}.

\begin{table*}[hptb]
\centering
\begin{tabular}{c c c c c c}
\hline
Sample & Catalog(s) & Area & IR-selected AGN & X-ray selected AGN & Number of AGN used for LF \\\hline\hline
X-ray (\S\ref{sec:xsample}) & XMM-SERVS & \SI{5.3}{deg^2} & --- & 5071 & 3427 \\
IR (\S\ref{sec:irsample}) & HELP & \SI{9.0}{deg^2} & 5918 & --- & 4268 \\
Comparative (\S\ref{sec:crossmatch}) & XMM-SERVS, HELP & \SI{3.4}{deg^2} & 958 & 1068 & 917 (IR), 829 (X-ray) \\
\hline
\end{tabular}
\caption{Summary of samples used to identify AGN and construct the LF. The data from XMM-SERVS \citep{chen_xmm-servs_2018} and HELP \citep{shirley_help_2019, shirley_help_2021} are used for the X-ray and IR samples, respectively, and both are used for the comparative sample. The last column shows the number of AGN used to construct the LFs presented in \S\ref{sec:results} after sample binning and incompleteness corrections have been applied. We note that the difference in number between selected AGN and the number used for the LF is mostly attributable to the the sample binning requiring $z > 0.5$ and $ L > \SI{2e43}{erg~s^{-1}} $ (see Fig.~\ref{fig:lum_vs_red} and \S\ref{sec:binning}).}
\label{tab:samples}
\end{table*}

\subsection{Constructing AGN samples}
\label{sec:selectir}
To avoid biasing the AGN LF, selection criteria need to provide an accurate and complete set of sources.
X-ray selection can be done using a luminosity cut because there are very few astrophysical objects that reach such high luminosities, but this selection will not identify heavily obscured AGN \citep{brandt2015A&ARv..23....1B, almeida_nuclear_2017}.
However, corrections for obscuration and the identification of obscured AGN are outside the scope of this paper.
In the IR, color distribution is often used to select AGN, but different criteria \citep{lacy_obscured_2004, donley_identifying_2012, stern_mid-infrared_2005} achieve different results. 
Our approach is to use a highly accurate X-ray AGN sample and a permissive color cut in the IR to give us a hopefully more complete but less accurate IR AGN sample.
We then use the XLF and X-ray sample to provide a check on our IRLF and IR sample (see \S\ref{sec:crossmatch} and \S\ref{sec:results} for more details).
Our samples are summarized in Table~\ref{tab:samples}.

\subsubsection{X-ray AGN sample}
\label{sec:xsample}
The XMM-SERVS catalog flags sources that are selected by different selection criteria (including a luminosity cut, X-ray to optical flux ratio, and spectroscopic identification).
A total number of $5071/5242 \approx 96.7\%$ of the sources are classified as AGN using these criteria. 

\subsubsection{IR AGN sample}
\label{sec:irsample}
The \cite{lacy_obscured_2004} selection criterion identifies the largest number of objects (compared to other IR color selection criteria) and so is effective at identifying a more complete sample of AGN.
With this selection we have $5918$ IR AGN accounting for $5918/18328 \approx 32\%$ of the total number of sources from the HELP catalog in the XMM-LSS field.
The distribution of AGN in IR color space is distinct from that of star-forming galaxies (SFGs) so we expect contamination of the IR fluxes due to SFGs to be relatively small \citep{sajina_simulating_2005, lacy_spitzer_2015}.
However, some non-AGN objects satisfy the criterion and so suffers in its accuracy.
We discuss incompleteness and our associated corrections in \S\ref{sec:inc_corr}.

\subsubsection{Comparative sample}
\label{sec:crossmatch}
In order to overlay the X-ray AGN onto the IRLF and vice versa, we use the same selection criteria on a subsample of sources from the HELP and XMM-SERVS catalogs.
We perform a crossmatch of the HELP and XMM-SERVS catalogs in a subregion of the XMM-LSS field where they overlap to get a sample of sources with both X-ray and IR fluxes.
The crossmatch uses \texttt{TOPCAT} \citep{taylor_topcat_2005} and a simple matching criterion that requires the distance between the sources to be less than $r' = 1"$ using the optical coordinates from XMM-SERVS and IRAC coordinates from HELP.
There are no multiple associations with this crossmatch.
We consider the crossmatch in a subregion of the XMM-LSS field with area \SI{\sim 3.38}{\deg^2}.
The HELP catalog is based on MIPS 24 micron selection where the random probability of a source falling with a square arcsecond is $\mathcal{O}(10^{-3})$ so the crossmatch is unlikely to have a false association.
Fig.~\ref{fig:xmatch_sample} shows the full X-ray and IR samples as well as the comparative region.
This sample, which we refer to as the \textit{comparative} sample, has 958 IR selected AGN and 1068 X-ray selected AGN.
\edit2{The discrepancy in the numbers of selected AGN between the two bands is to be expected because different selection criteria are used to identify the sources as AGN.}

\subsection{AGN luminosities}
\label{sec:lum}
The luminosity of the AGN is computed for the given fluxes and redshifts as
\begin{equation}
L = 4\pi d^2_L(z)F,
\end{equation}
where $d_L$ is the luminosity distance.
% In terms of the flux density, the luminosity is 
% \begin{equation}
%     L = 4\pi d^2_L(z)\nu F_\nu.
% \end{equation}

\begin{figure}[hptb]
\centering
\includegraphics[width=0.95\columnwidth]{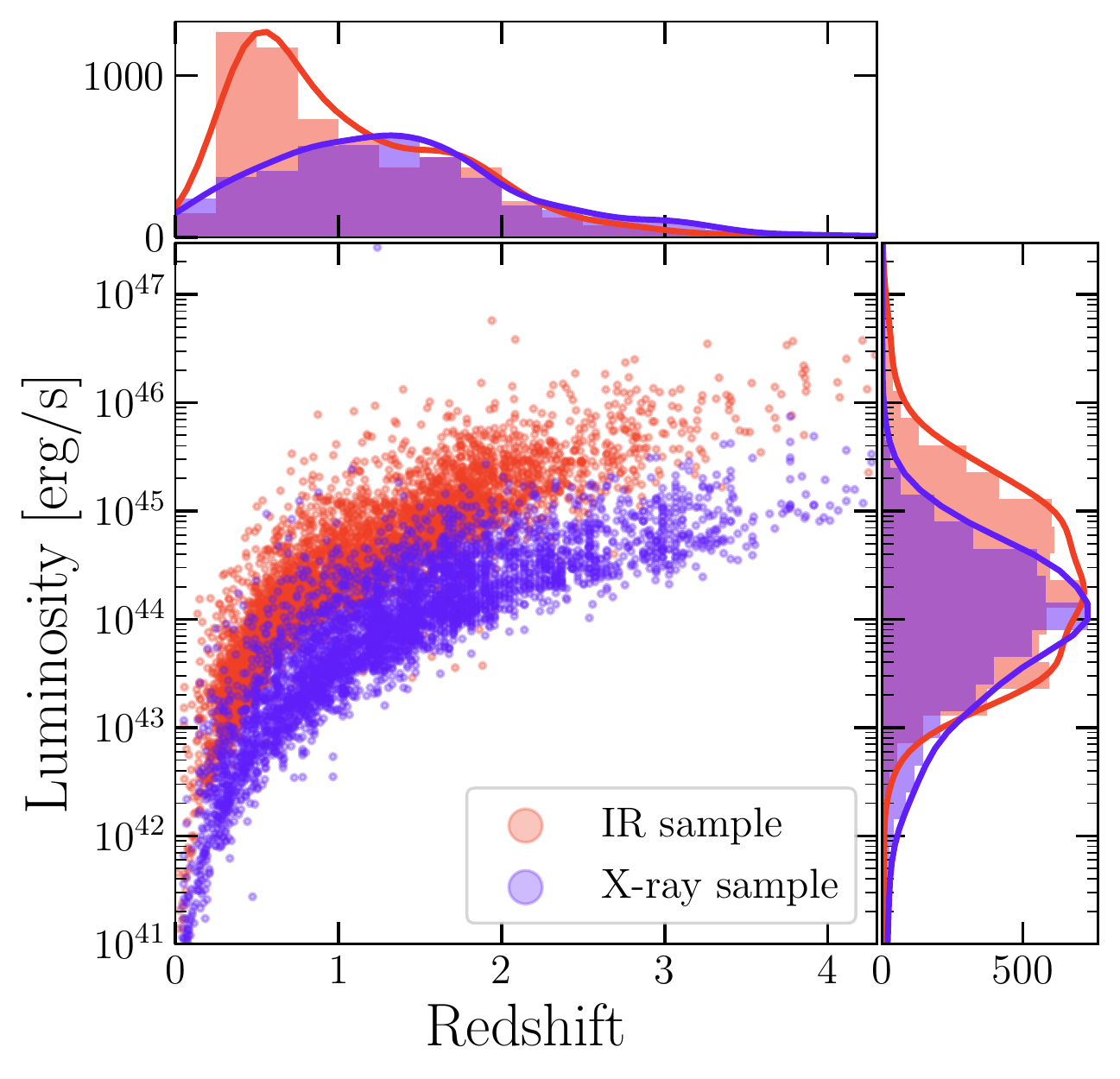}
\caption{Luminosity-redshift diagram of the IR (red) and X-ray (blue) selected AGN samples from the HELP and XMM-SERVS catalogs in red and purple, respectively. The histograms show the samples binned in redshift and luminosity with an unnormalized KDE overlaid to approximate the true distribution of the samples. Spectroscopic redshifts are chosen preferentially over photometric redshifts. The luminosities include K-corrections but no bolometric corrections.}
\label{fig:lum_vs_red}
\end{figure}

We restrict our sample to the sub-set of objects with available 8 micron IRAC photometry and 24 micron MIPS photometry in order to calculate the K-correction.
This restriction may introduce some incompleteness, however we attempt to address this in \S\ref{sec:inc_corr}.
For simplicity, the spectrum is assumed linear between the \SI{5}{\micro m} and \SI{8}{\micro m} bands and between the \SI{8}{\micro m} and \SI{24}{\micro m} bands. 
%\jr{EH: this is somewhat redshift dependent}
Then the correction is the interpolated value between these two data points.

For the X-ray sample, we similarly use the interpolated spectrum between the soft and hard bands to determine the spectral index.
We find a mean spectral index of $\Gamma\approx 1$ for our AGN.

\begin{figure}[hptb]
\centering
\includegraphics[width=\columnwidth]{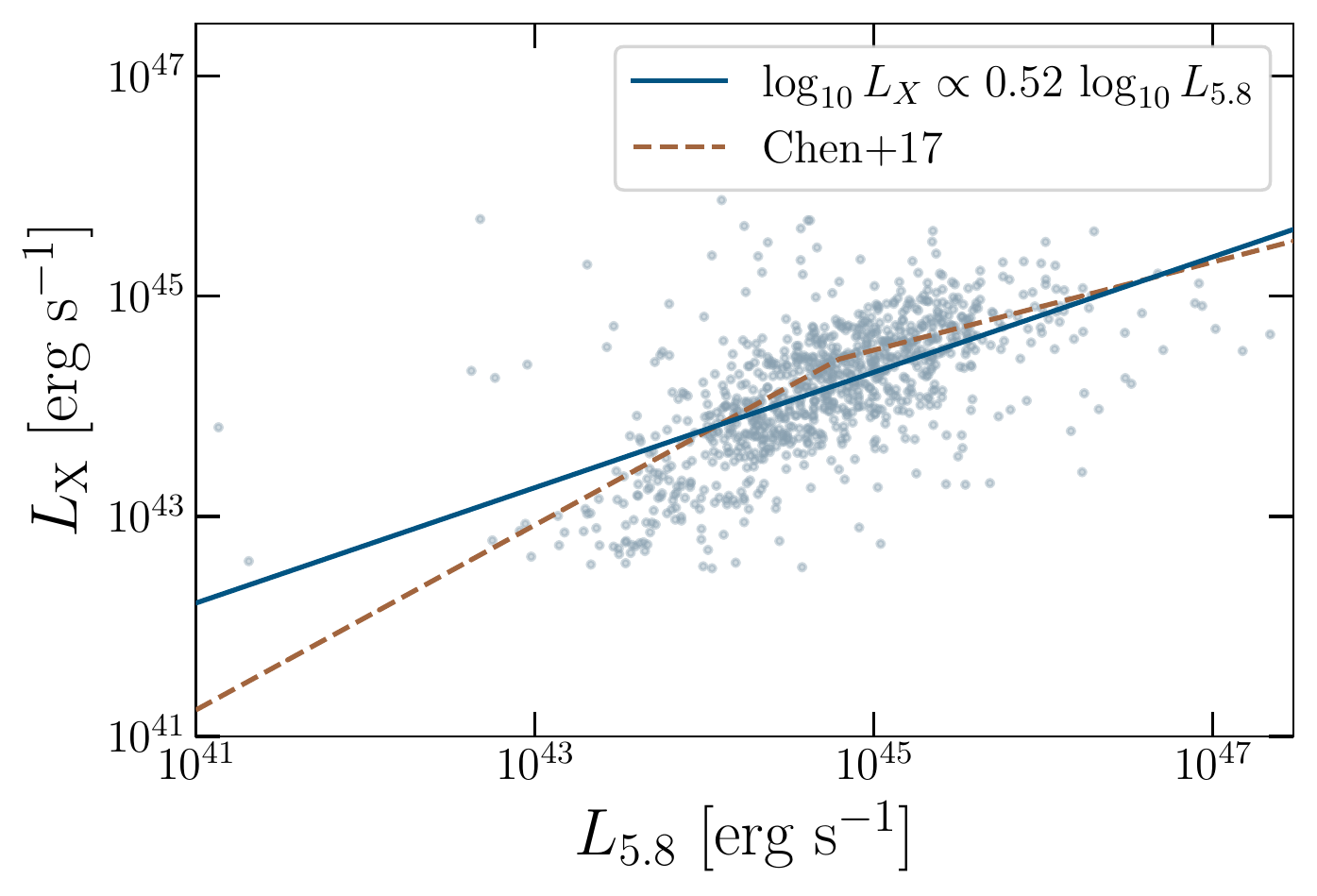}
\caption{IR and X-ray luminosities of objects selected as AGN in both X-ray and IR. The blue line is a simple power law fit to the data. We include a comparison to \cite{chen_x-ray_2017}.}
\label{fig:lvsl}
\end{figure}

With the K-correction computed as a spectral index $\Gamma$, the luminosity is given by
\begin{equation}
L = 4\pi d_L^2(z) F \times (1 + z)^{-1 + \Gamma}.
\end{equation}
In Fig.~\ref{fig:lum_vs_red}, we show the luminosity-redshift diagram of our X-ray selected and MIR selected AGN.
In Fig.~\ref{fig:lvsl}, we show the luminosities in IR and X-ray for our comparative sample as well as a power law fit to the data.

% \subsubsection{SEDs}
% Plots of SEDs for all AGN in X-ray and IR in Fig.~\ref{fig:seds}.

% \begin{figure*}[]
%     \centering
%     \gridline{\fig{figures/sed_ir.pdf}{\columnwidth}{IR}\fig{figures/sed_xray.pdf}{\columnwidth}{X-ray}}
%     \caption{SEDs of all selected AGN in both X-ray and IR. In the IR sample, the long tail in the faint \SI{24}{\micro m} band corresponds to lower luminosity AGN. I don't have a good explanation for why this is true!}
%     \label{fig:seds}
% \end{figure*}

% \begin{figure}
%     \centering
%     \includegraphics[width=\columnwidth]{figures/ir_seds.pdf}
%     \caption{Violin plots of HELP AGN SEDs. 
%     The violin plots are truncated to the region defined by the sample.
%     From this plot, it is clear that the problematic (negative) K-corrections stem from the lowest luminosity bins for each redshift bin.}
%     \label{fig:seds}
% \end{figure}

\section{Methods: constructing the luminosity function}
\label{sec:methods}
The value of the luminosity function at a given luminosity and redshift is the number of objects per co-moving volume per luminosity interval 
\begin{equation}
\label{eq:lf}
\phi(L, z) = \frac{d\Phi}{dL} = \frac{d^2N(L, z)}{dVdL}.
\end{equation}
It has units of \si{Mpc^{-3}~erg^{-1}~s}.
Often, the quantity $d\Phi/d\log L$ is used in which case the units are \si{Mpc^{-3}} (i.e. true number density).

\subsection[1/Vmax method]{\(1/V_{\mathrm{max}}\) method}
\label{sect:vmax}
The $1/V_\mathrm{max}$ method yields an unbiased, binned, non-parametric estimate of the LF.\footnote{\cite{page_improved_2000} provide an alternative approach that improves estimates near the flux limit of the survey and estimate of uncertainties in each bin. However, we use the $1/V_\mathrm{max}$ method for simplicity.}
This method exploits the fact that brighter objects can be seen within a larger co-moving volume $V_\mathrm{max}$ than fainter objects but are also identified more frequently.
Each source contributes to the number density; brighter objects can be identified in a larger region and so naturally make a smaller contribution to the value of the LF.

To build the LF $\phi$ using the $1/V_{\mathrm{max}}$ method, first we calculate the maximum effective volume that each object could occupy for the given survey
\begin{equation}
V_{\mathrm{max}}=\int d\Omega\int_{L_\mathrm{min}}^{L_\mathrm{max}}dL \int^{z_{\mathrm{max}}}_{z_{\mathrm{min}}} \frac{dV_c}{dzd\Omega}dz
\end{equation}
where
\begin{equation}
\label{eq:comov_vol}
\frac{dV_c}{dzd\Omega} = \frac{c}{H_0}\frac{d_L^2(z)}{(1+z)^2 \sqrt{\Omega_m(1+z)^3+\Omega_\Lambda}}
\end{equation}
is the differential comoving volume element and where $A=\int d\Omega$ is the area of the survey. 

% \begin{figure}[b]
%     \centering
%     \includegraphics[width=\columnwidth]{figures/ir_k_corrections.pdf}
%     \caption{K-corrections for the IR data. 
%     Some objects exhibit a negative K-correction which seems uncharacteristic of most AGN spectra (?). 
%     These objects are not an appreciable fraction of the full sample so I haven't tried to address their effect on the LF. 
%     However, we may want to say somehting about them...}
%     \label{fig:k_corrs}
% \end{figure}

The bounds of the integral over $z$ are the maximum and minimum redshift within which an object of luminosity $L$ could be detected.
For the binned LF, $z_{\rm min}$ is given by the lower edge of the redshift bin and $z_{\rm max}$ is given by the upper edge of the redshift bin unless the object would fall below the luminosity limit in a given bin, then $z_{\rm max}$ is given by the luminosity corresponding to the flux limit.
The bounds of the integral over $L$ are the maximum and minimum luminosities the object could have in that luminosity bin (i.e. the bin edges). 
The integral over $\Omega$ is over the field area. 
(N.B.: These limits directly depend on flux limits of the instrument/survey.)
Note that with this definition, the units of $V_\mathrm{max}$ are \si{Mpc^3~erg~s^{-1}}.

Using these $V_\mathrm{max}$ values, we can obtain an unbiased estimate of the LF as
\begin{equation}
\phi(L) = \sum_{i=1}^{N}\frac{1}{V_{\mathrm{max},\,i}}.
\end{equation}

To quantify the uncertainty in the binned estimate of our LF, we include Poisson errors based on the number of objects in a given redshift and luminosity bin.
We use the corrections tabulated in \cite{gehrels_confidence_1986} for upper and lower limits based on Poisson statistics. 
Uncertainties due to photometric redshifts are discussed in \S\ref{sec:mc}. 

% The errors for these estimates are 
% \begin{equation}
%     \sigma=\sqrt{\sum_i \frac{1}{V_{\mathrm{max},\,i}^2}}.
% \end{equation}
% In practice, however, we will obtain uncertainties on our LF values using Monte Carlo methods.

% \begin{figure}[hptb]
% \centering
% \includegraphics[width=0.95\columnwidth]{figures/venn_diagrams_redshifts_corrected.pdf}
% \caption{Venn diagram of AGN in each redshift bin. The IR count (red) corresponds to the number of IR-selected AGN selected from HELP in the subregion of the comparative sample and the X-ray count (blue) corresponds to X-ray selected AGN from the XMM-SERVS catalog in the subregion of the comparative sample. The overlap is the number of AGN selected in both X-ray and IR in the comparative subfield.}
% \label{fig:venn}
% \end{figure}

\subsection{Sample binning}
\label{sec:binning}

Because the $1/V_{\rm max}$ method is a binned estimate of the LF, we need to bin our sample in $L-z$ space such that each bin has a sufficient number of objects.
We use logarithmically spaced bins between $0.5 \leq z \leq 4.0$ and \SI{2e43}{erg~s^{-1}} $\leq L \leq$ \SI{3e46}{erg~s^{-1}} for both the XLF and IRLF.
Excluding the highest redshift bin, there is at least one luminosity bin in each redshift bin that has more than 100 AGN.
In the last redshift bin ($3.17\leq z <4.0$), the highest bin count is 38.
In most redshift bins, the large number of sources and $\sigma_\mathrm{NMAD} < 0.1$ are sufficient to avoid significant bias due to errant photometric redshift estimates (see \S\ref{sec:binning} for more details).
Fig.~\ref{fig:lum_vs_red} shows the unbinned distribution in the $L-z$ plane.
% Also, Fig.~\ref{fig:venn} shows the number of AGN in each redshift bin.

% We show the breakdown of our sample in our redshift bins in both X-ray and IR in Fig.~\ref{fig:venn}.

\subsection{Incompleteness corrections}
\label{sec:inc_corr}

\begin{figure*}[hptb]
\centering
\gridline{\fig{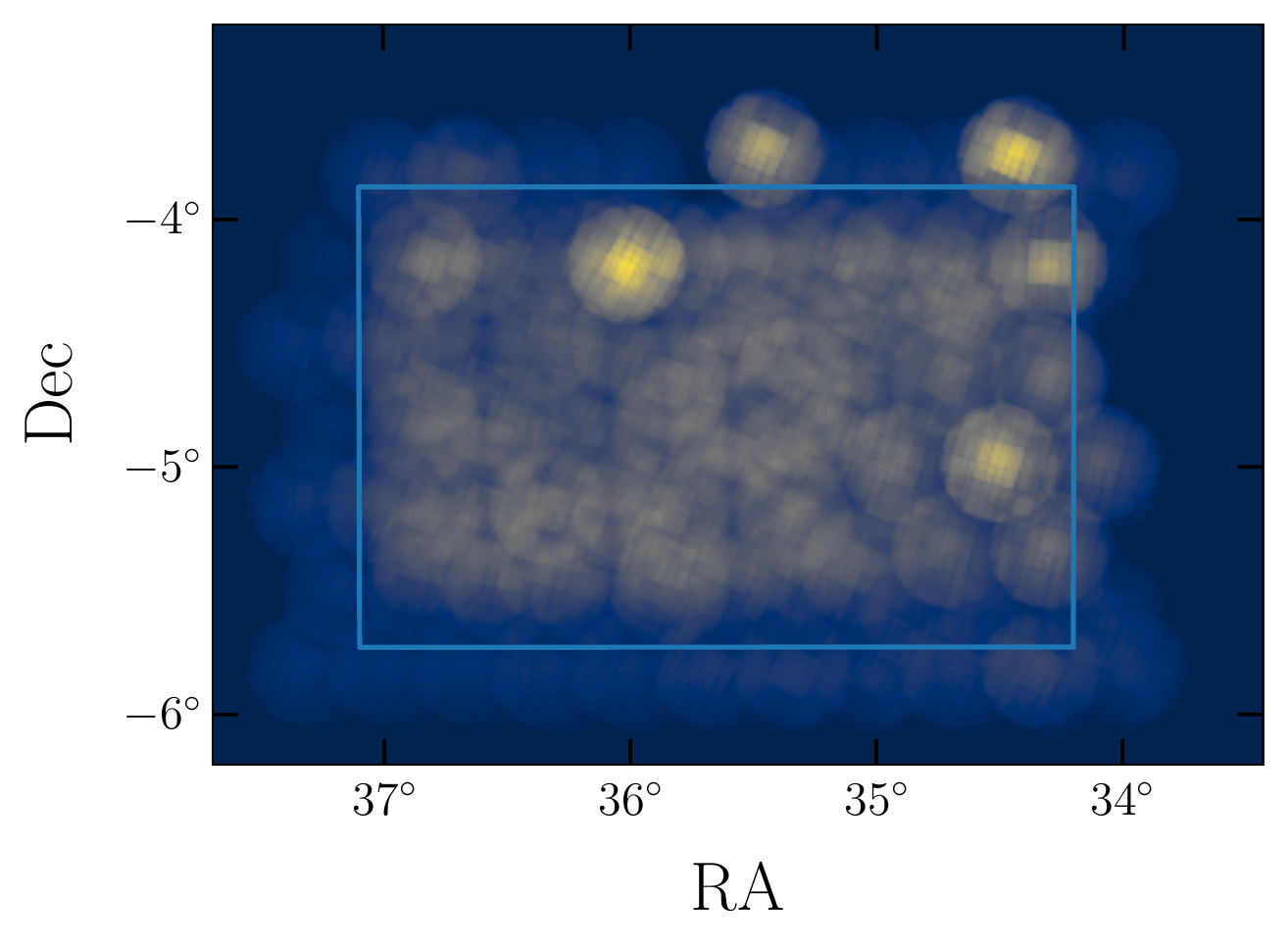}{\columnwidth}{}\fig{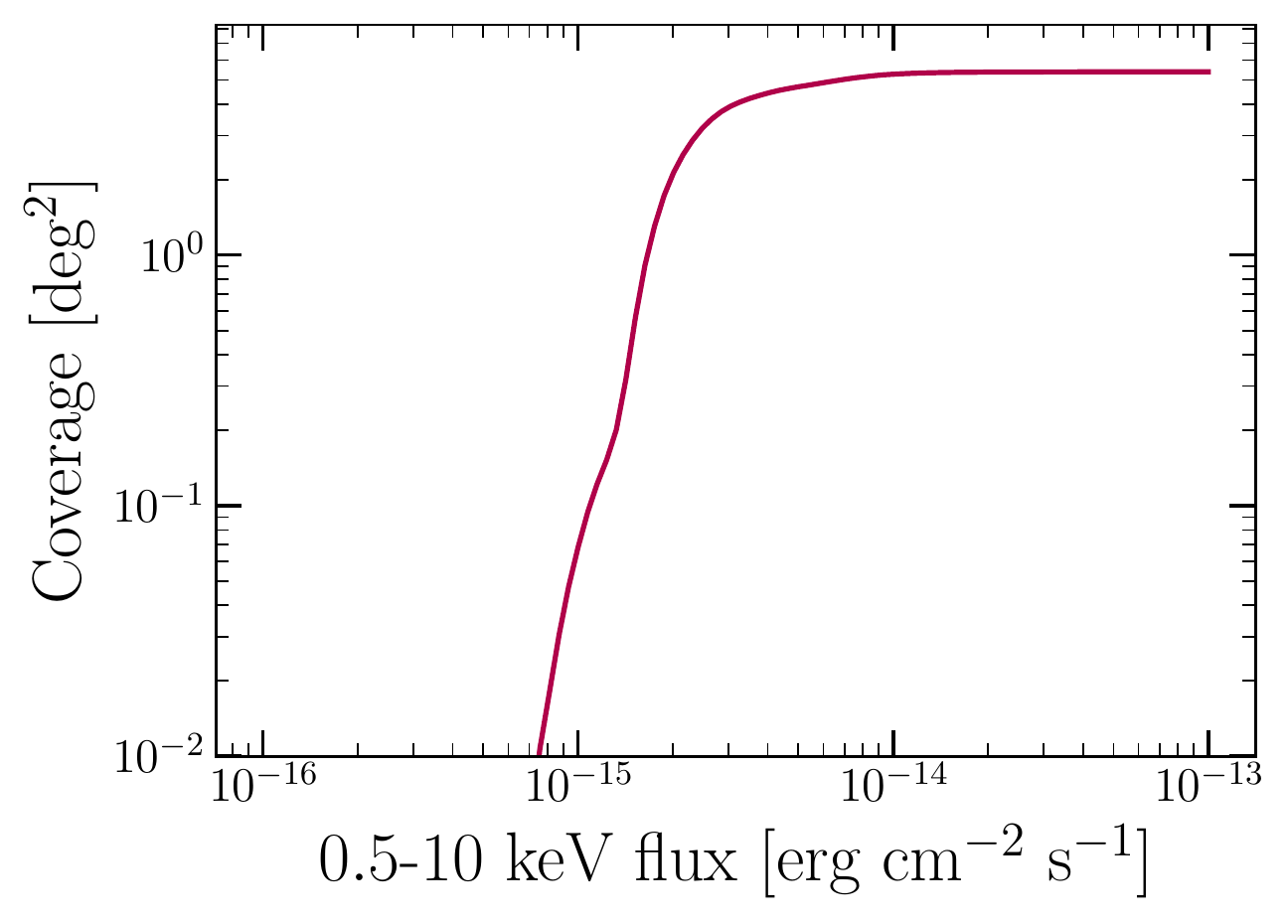}{\columnwidth}{}}
\caption{\textit{Left panel:} Exposure map of \textit{XMM-Newton} over the XMM-LSS field. The blue box is the region used to generate the XMM-SERVS catalog. \edit2{\textit{Right panel:}} X-ray coverage as a function of flux for XMM-SERVS catalog. Note that the coverage is uncorrected for Eddington bias.}
\label{fig:exposure_map}
\end{figure*}

In principle, the coverage of the survey field may not be uniform as a function of flux.
The coverage of SWIRE over the XMM-LSS field is roughly constant.
However, the X-ray coverage is flux-dependent. 
In order to account for this effect, we modify the area in Eq.~\ref{eq:comov_vol}, letting $\Omega=\Omega(F)$ where $F = L/4\pi d_L^2(z)$ using the exposure map to calculate the coverage curve shown in Fig.~\ref{fig:exposure_map}. 

Additionally, we add a correction for incompleteness of our AGN sample.
Obscuring material around luminous AGN can dramatically affect the observed spectrum.
In particular, Compton-thick obscuration reduces the reliability and completeness of any X-ray selected AGN sample. 
Galaxies undergoing high rates of star formation may confuse AGN selection attempts in IR bands.
Of course, less rapidly accreting AGN will be intrinsically fainter and thus may also be missed.
To correct for the incompleteness of our AGN sample, we follow the prescription in \cite{ranalli_210_2016} and assume the incompleteness correction is only a function of flux $F$. 
We define the correction factor as the ratio between the number of selected AGN to the total number of objects as a function of flux
\begin{equation}
\label{eq:inc_corr}
C(F) = \frac{N_\mathrm{AGN}(F)}{N_\mathrm{tot}(F)}.
\end{equation}
This correction is used to calculate an ``effective coverage" for each object
\begin{equation}
\Omega_\mathrm{eff} (F) = C(F)\Omega(F).
\end{equation}
Care must be taken when choosing what we mean by the ``total'' sample in the denominator of Eq.\ref{eq:inc_corr}.
We take the approach that the ``total'' sample is the sample detailed in \S\ref{sec:sample} before selection criteria are applied.

For the comparative sample, we use the XMM-SERVS coverage because it is the limiting factor in the crossmatch.
We take the ``total'' sample to be all the objects in the comparative region with a redshift.

\begin{figure}[hptb]
\centering
\includegraphics[width=\columnwidth]{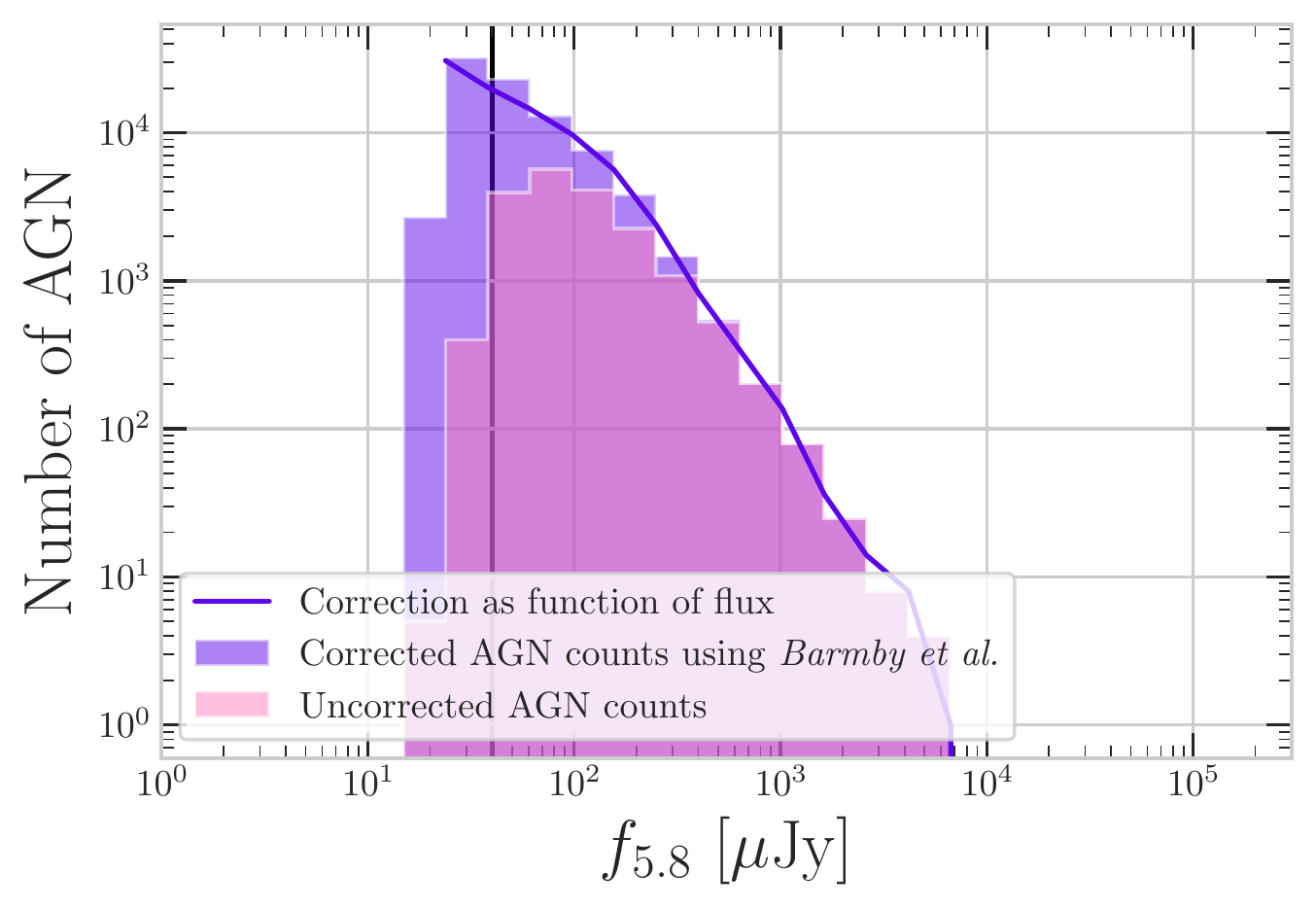}
\caption{Incompleteness corrections using incompleteness estimates from \cite{barmby_catalog_2008}.
The pink histogram shows the number density of AGN in our sample as a function of flux.
The purple histogram includes the incompleteness corrections. We impose a flux density cut at \SI{40}{\micro Jy} (indicated by the vertical black line) to avoid the large uncertainties associated with significant incompleteness.}
\label{fig:corr_flux}
\end{figure}

To more robustly address the incompleteness of our AGN sample in the IR band, we include corrections based on the results of \cite{barmby_catalog_2008} in the EGS across all four IRAC bands.
In their analysis, \cite{barmby_catalog_2008} inject artificial sources into the IRAC mosaics and derive completeness estimates based on their success in recovering these artificial sources.
The validity of using these corrections a comparison of counts in these fields is discussed in \cite{lacy_spitzer_2021}.

In order to account for differing coverage and field area, we assume that at higher fluxes both our sample and the \cite{barmby_catalog_2008} sample are 100\% complete.
At these high fluxes, we match the \cite{barmby_catalog_2008} counts to ours and use the ratio of their counts to ours to correct for incompleteness at lower fluxes.
These ratios are interpolated and then used to correct each source for incompleteness as a function of its flux.
% \jr{Say something like: the detection limits are roughly the same so...50\% completeness limit of EGS field is \SI{1.5 }{\micro Jy} at \SI{3.6}{\micro m}.\citep{barmby_catalog_2008} and it is about \SI{1.9}{\micro Jy} at \SI{3.6}{\micro m} for the XMM-LSS field \citep{mauduit_spitzer_2012}}
Fig.~\ref{fig:corr_flux} shows the effect of these corrections on the number of fainter sources.
The purple line shows the corrected number of counts using the interpolated function in good agreement with the corrected bin counts.
In the lowest flux bins, the completeness is less than 1\%.
The high levels of incompleteness at these lower fluxes may affect the reliability of our conclusions and introduce uncertainty.
To avoid this, we introduce a flux cut at \SI{40}{\micro Jy}.
The result of this flux cut and the sample binning discussed in the previous section yield the number of AGN used to construct the LF (shown in Table~\ref{tab:samples}).

\subsection{Using Monte Carlo to account for redshift uncertainty}
\label{sec:mc}
Thus far our analysis has not taken into account the effects of uncertainties in the photometric redshifts.
The SED fitting procedure (for HELP, \cite{duncanI2018MNRAS.473.2655D} \& \cite{duncanII2018MNRAS.477.5177D}; for XMM-SERVS, \cite{chen_active_2021}) attempts to estimate the redshift of an AGN by fitting templates to the observed fluxes for each object.
While this can provide a good redshift estimate for some objects, incorrect templates or insufficient photometric data can lead to catastrophic failure (i.e. where $\Delta z/z_\mathrm{spec}>0.1$) in redshift estimation.
Both the IR and X-ray samples include objects with photometric redshifts.

\begin{figure*}[hptb]
\centering
\includegraphics[width=0.8\textwidth]{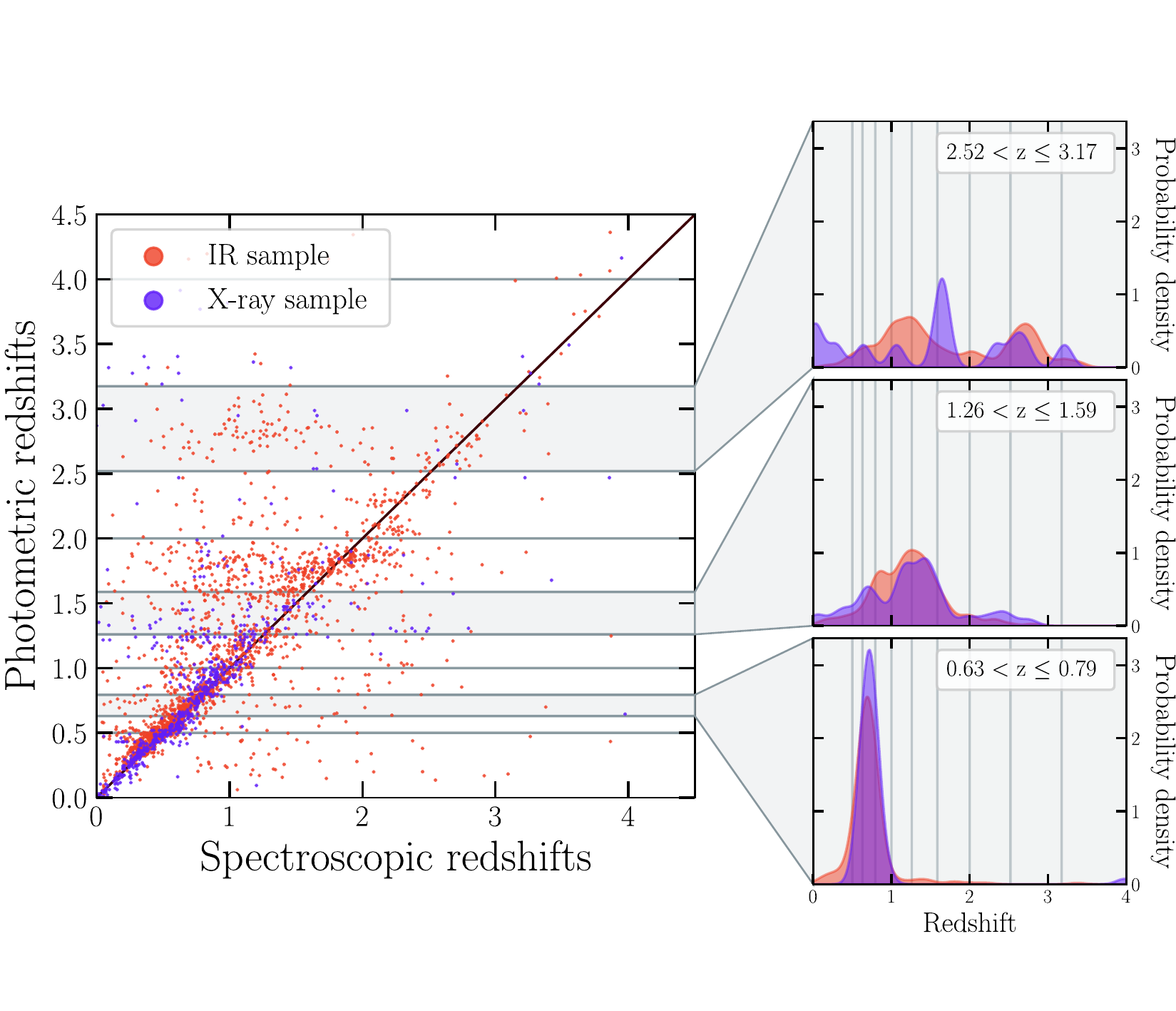}
\caption{Comparison of photometric and spectroscopic redshifts for the X-ray and IR samples. The redshift bins are overlaid along the photometric redshift axis. In many redshift bins, there is a high rate of catastrophic failure in the photometric redshift SED-fitting routine. 
We aim to address this source of error by constructing a statistical description of the catastrophic failures within each redshift bin. \textit{Right inset}: The KDEs of redshift for objects with both photometric and  spectroscopic redshifts. These PDFs show the scatter of the spectroscopic redshifts for objects that have a photometric redshift in that redshift bin. The vertical lines indicate the redshift bin edges indication how likely an AGN with a given photometric redshift is to have a `true' redshift outside of that bin. When performing the Monte Carlo, each photometric redshift is replaced by a random pull from the appropriate KDE.}
\label{fig:z_v_z}
\end{figure*}

Our goal is to get as good an estimate as possible of the `true' (i.e. spectroscopic) redshift of an object with a given photometric redshift.
To this aim, we construct the PDF of the `true' redshifts that fully captures the uncertainties and catastrophic failures of the SED fitting routine.
We empirically construct this PDF for each redshift bin using the subsample of AGN that have both photometric and spectroscopic redshifts.
This procedure allows us to model the uncertainties for the photometric redshifts for the full sample of AGN based on the subsample used to construct the PDF.

This analysis assumes that the subsample of AGN with both photometric and spectroscopic redshifts is representative of the full sample; however, this is not necessarily true.
Objects with spectroscopic redshifts tend to be bright and strong-lined which may introduce a bias into our analysis.
We are primarily concerned with the construction of the AGN LF so we use the LF to probe the effects of this bias.
In \S\ref{sec:results}, we compare the LF using the full AGN sample and the sample of AGN with spectroscopic redshifts.
Additionally, to further understand the effects of our methods, we use simple Gaussian uncertainties on the photometric redshifts and compare to our main results in Appendix~\ref{app:gauss}.

Rather than treat the objects in a bin individually, we treat the objects statistically by making a PDF that encodes the frequency and degree of catastrophic failure for a given photometric redshift.
With these PDFs, we perform a Monte Carlo, drawing from the appropriate PDF for each object that lacks a spectroscopic redshift.
The comparison of photometric and spectroscopic redshifts is shown in Fig.~\ref{fig:z_v_z}.
In each redshift bin (i.e. horizontal band in the figure), the scatter of the `true' redshift for a given photometric redshift is shown.
% For example, for a photometric redshift of $z=1.75$, there are objects with spectroscopic redshifts less than $z=0.5$ and more than $z=3$.

In practice, the kernel density estimation routine from \texttt{scikit-learn}, the machine-learning library for Python, with a bandwidth of 0.1 is used to construct an estimate of the PDF in each redshift bin.
The bandwidth value of 0.1 is chosen to roughly represent an uncertainty of a few percent in the spectroscopic redshifts.
However, we found that the LF is insensitive to modest changes in the bandwidth.
The insets in Fig.~\ref{fig:z_v_z} show the PDFs for three of the redshift bins in both X-rays and IR.
% \jr{make plot of r band magnitude for objects with spectroscopic redshifts vs without}

\section{Results}
\label{sec:results}
Our main results are the IRLFs in Fig.~\ref{fig:ir_lf_specz} and Fig.~\ref{fig:ir_lf_xmatch} and the XLFs in Fig.~\ref{fig:xray_lf_specz} and Fig.~\ref{fig:xray_lf_xmatch}.
We find that our LFs are in good agreement with parametric models from the literature \citep{lacy_spitzer_2015, aird_evolution_2010, ranalli_210_2016} as discussed below and our analysis is able to reproduce consistent LFs while varying selection criteria and subsampling based on redshift quality.

\subsection{IRLF}
\begin{figure*}[hptb]
\centering
\includegraphics[width=\textwidth]{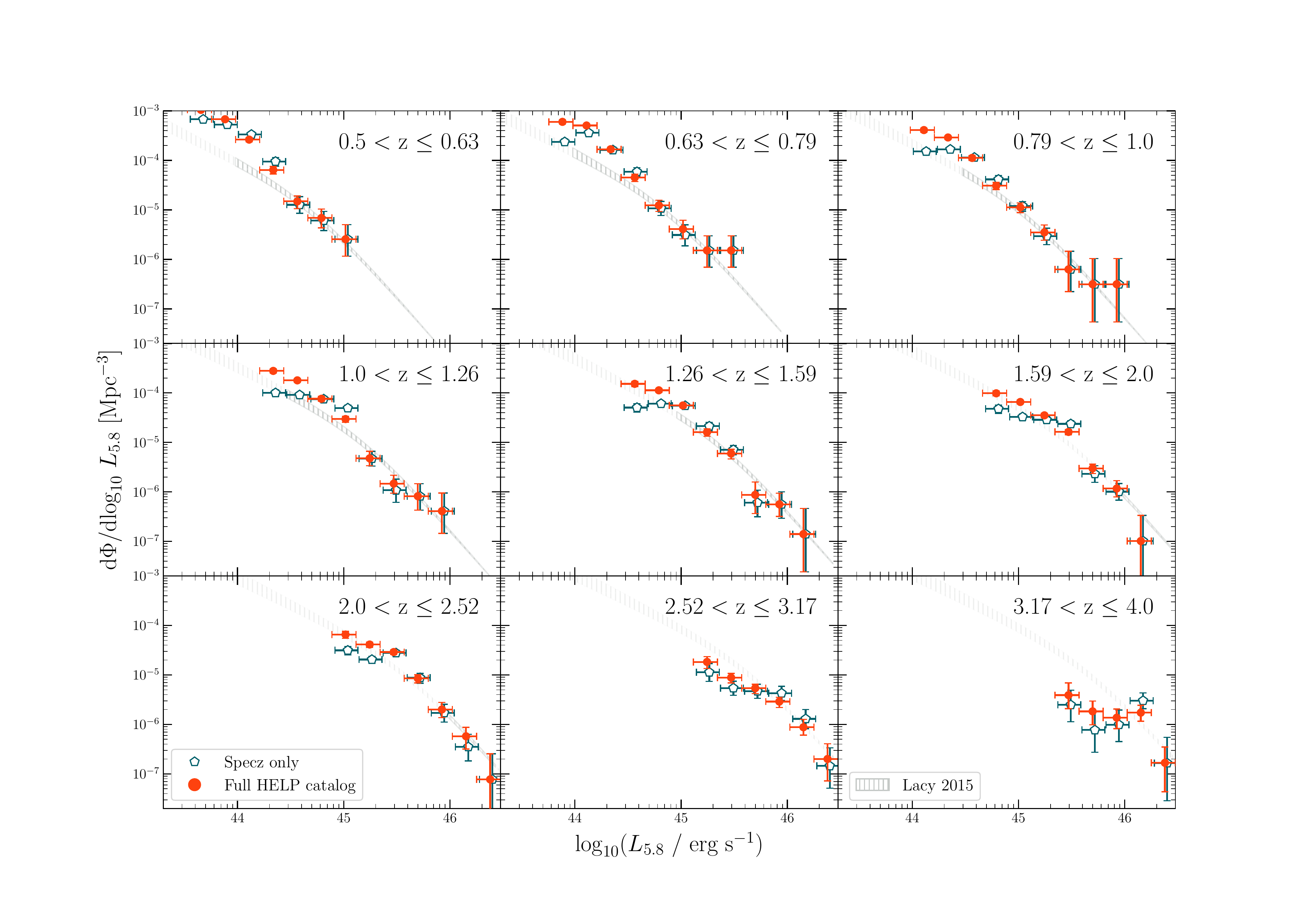}
\caption{IRLF constructed using the HELP catalog with uncertainties estimated by MC of N=10000. The IRLF constructed using only spectroscopic redshifts is also shown. The 68\% confidence interval of the \cite{lacy_spitzer_2015} parametric IRLF is shown for comparison--the lighter colored, hatched region shows where \cite{lacy_spitzer_2015} extrapolated their model; darker grey indicates the redshift and luminosity intervals supported by the data to construct their LF;. Our sample reaches higher redshifts and lower luminosities than the \cite{lacy_spitzer_2015} sample.  We note good agreement with these previous results. However, at higher redshift there is indication of a low-end luminosity flattening of the LF. This is not in contradiction of the \cite{lacy_spitzer_2015} results because their sample did not include such luminous AGN at high redshift (i.e. their  model has been  extrapolated to these redshifts).}
\label{fig:ir_lf_specz}
\end{figure*}

In Fig.~\ref{fig:ir_lf_specz}, we present the IRLF for the HELP sample as well as the subsample of IR-selected AGN with spectroscopic redshifts (`spec-z only').
For the spectroscopic redshift sample, we adjust the \cite{barmby_catalog_2008} bright-end counts to match the subsample as in Fig.~\ref{fig:corr_flux}.
The spectroscopic redshift sample also includes the incompleteness correction defined in Eq.~\ref{eq:inc_corr}.
\cite{lacy_spitzer_2015} fit a pure power law evolution model to their data.
Both LFs show good agreement with the evolution model in \cite{lacy_spitzer_2015}.
However, we note that there is a discrepancy between the full sample and the spectroscopic redshift only sample at low luminosities (see \S\ref{sec:discussion} for discussion of this discrepancy).

\begin{figure*}[hptb]
\centering
\includegraphics[width=\textwidth]{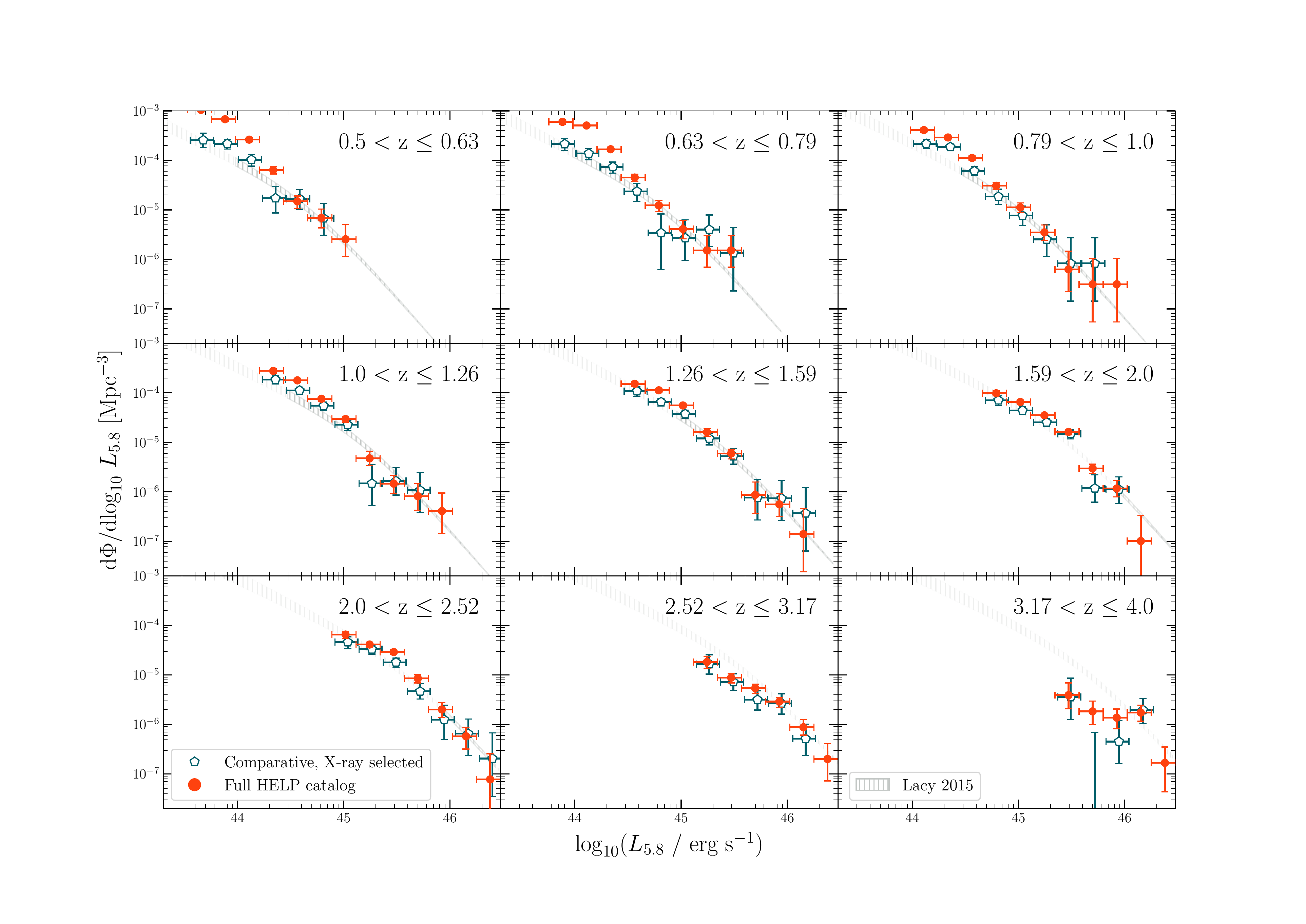}
\caption{IRLF as described in Fig.~\ref{fig:ir_lf_specz}. Here we plot the LF constructed using the comparative sample of X-ray selected AGN in comparison to the full AGN sample.}
\label{fig:ir_lf_xmatch}
\end{figure*}

In Fig.~\ref{fig:ir_lf_xmatch}, we show the same IRLF but include a comparison to the IRLF constructed in the comparative subfield with X-ray selected AGN.
The comparative LF includes the flux-dependent coverage factor described in \S\ref{sec:inc_corr} and shown in Fig.~\ref{fig:exposure_map} to account for the flux limits from XMM-SERVS that affect the IR sample after crossmatching.
Again, we find good agreement between these two LFs and \cite{lacy_spitzer_2015} model.

\subsection{XLF}
\begin{figure*}[hptb]
\centering
\includegraphics[width=\textwidth]{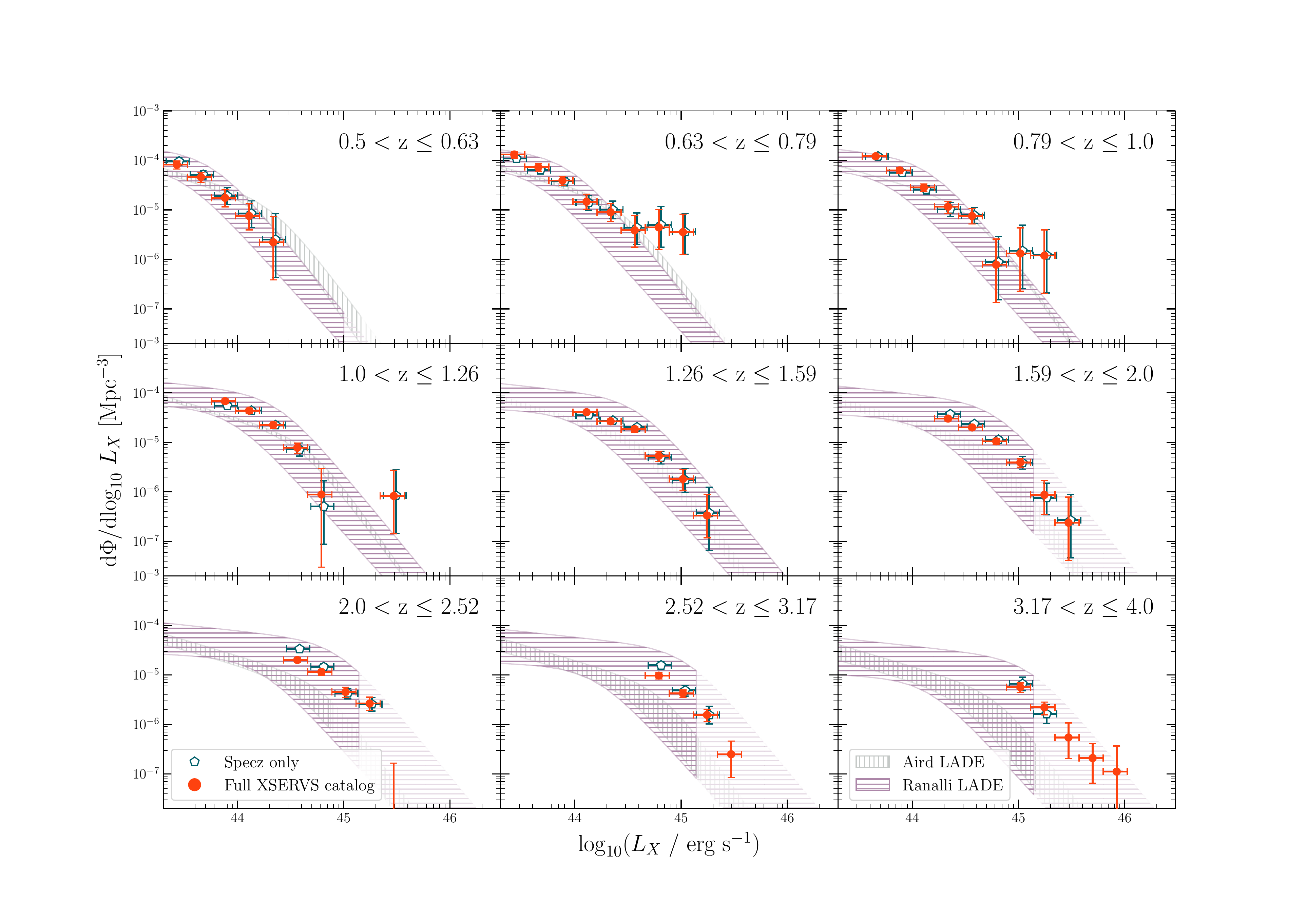}
\caption{XLF constructed using the XMM-SERVS catalog with uncertainties estimated by MC of N=10000. The XLF constructed using only spectroscopic redshifts is also shown. Also shown are the LADE evolution models of \cite{ranalli_210_2016, aird_evolution_2010}. The 68\% confidence intervals are hatched. The darker colored regions are where the models are supported by the data and lighter colored regions are where their models are extrapolated. In the higher redshift bin, our sample reaches higher luminosities than \cite{ranalli_210_2016} and \cite{aird_evolution_2010}. Our results are in good agreement with their models. In particular, our LF is in almost entirely within the confidence bands of the \cite{ranalli_210_2016} model.}
\label{fig:xray_lf_specz}
\end{figure*}

In Fig.~\ref{fig:xray_lf_specz}, we present the XLF for the full XMM-SERVS sample and the subsample of X-ray selected AGN with spectroscopic redshifts.
As with the IRLF, both XLFs are constructed using the same analysis outlined in \S\ref{sec:methods} but no Monte Carlo is performed for the spectroscopic redshift subsample.
Our LFs are in good agreement with the luminosity and density evolution (LADE) models from \cite{ranalli_210_2016} and \cite{aird_evolution_2010} which allow the overall amplitude and the location of the `knee' to evolve over redshift.
We note that since \cite{aird_evolution_2010} construct their XLF in the hard band, we correct their luminosities by a factor of $L_{X;\mathrm{full}}/L_{X;\mathrm{hard}}$, which is $\sim 1.04$ for our X-ray AGN sample.
\edit2{This value is derived from the average interpolated spectral index (see \S\ref{sec:lum}).}

\begin{figure*}[hptb]
\centering
\includegraphics[width=\textwidth]{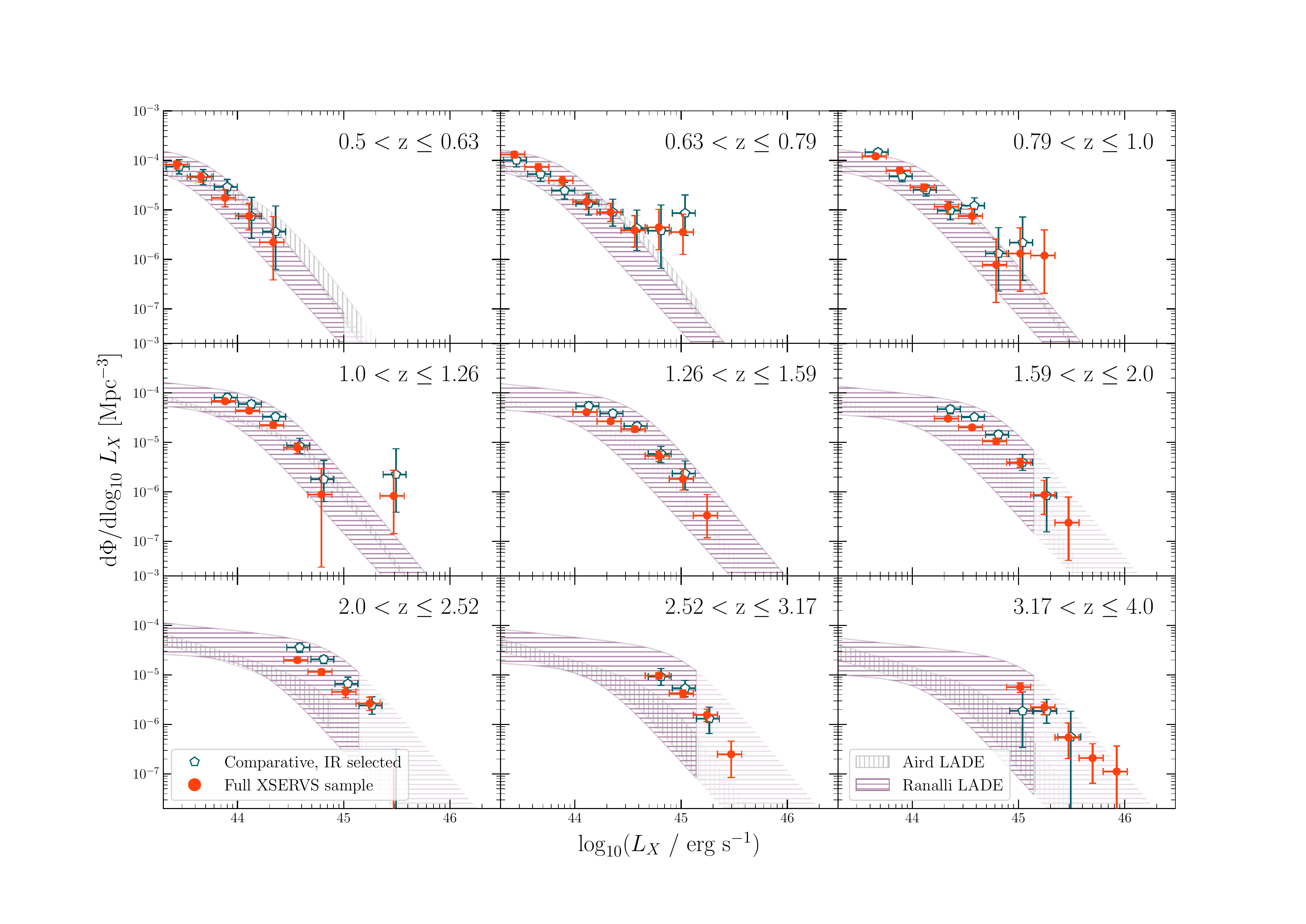}
\caption{XLF as shown in Fig.~\ref{fig:xray_lf_specz}. Additionally, we plot the LF constructed using the comparative sample of IR selected AGN.}
\label{fig:xray_lf_xmatch}
\end{figure*}

In Fig.~\ref{fig:xray_lf_xmatch}, the XLF of the IR selected AGN in the comparative subfield is shown.
Our XLFs are again in good agreement with each other and \cite{ranalli_210_2016} and \cite{aird_evolution_2010}.

% \jr{mention slight discrepancy at low l}

% \subsection{Comparing the IRLF \& XLF}
% \jr{Do we want to do this?}
\subsection{Potential X-ray clusters}
\begin{figure}[hptb]
\centering
\gridline{\fig{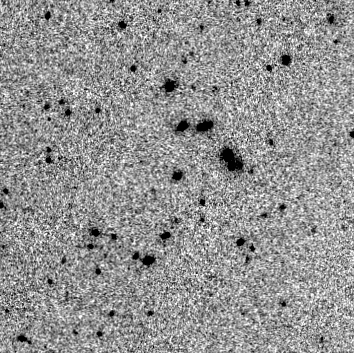}{0.15\textwidth}{}\fig{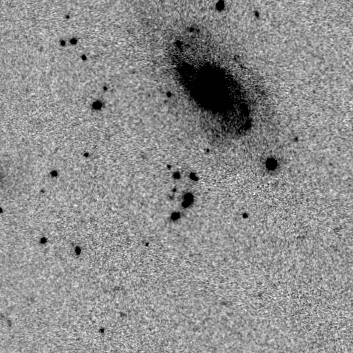}{0.15\textwidth}{}\fig{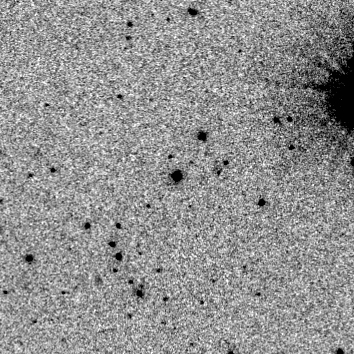}{0.15\textwidth}{}}
\vspace{-0.9 cm}
\gridline{\fig{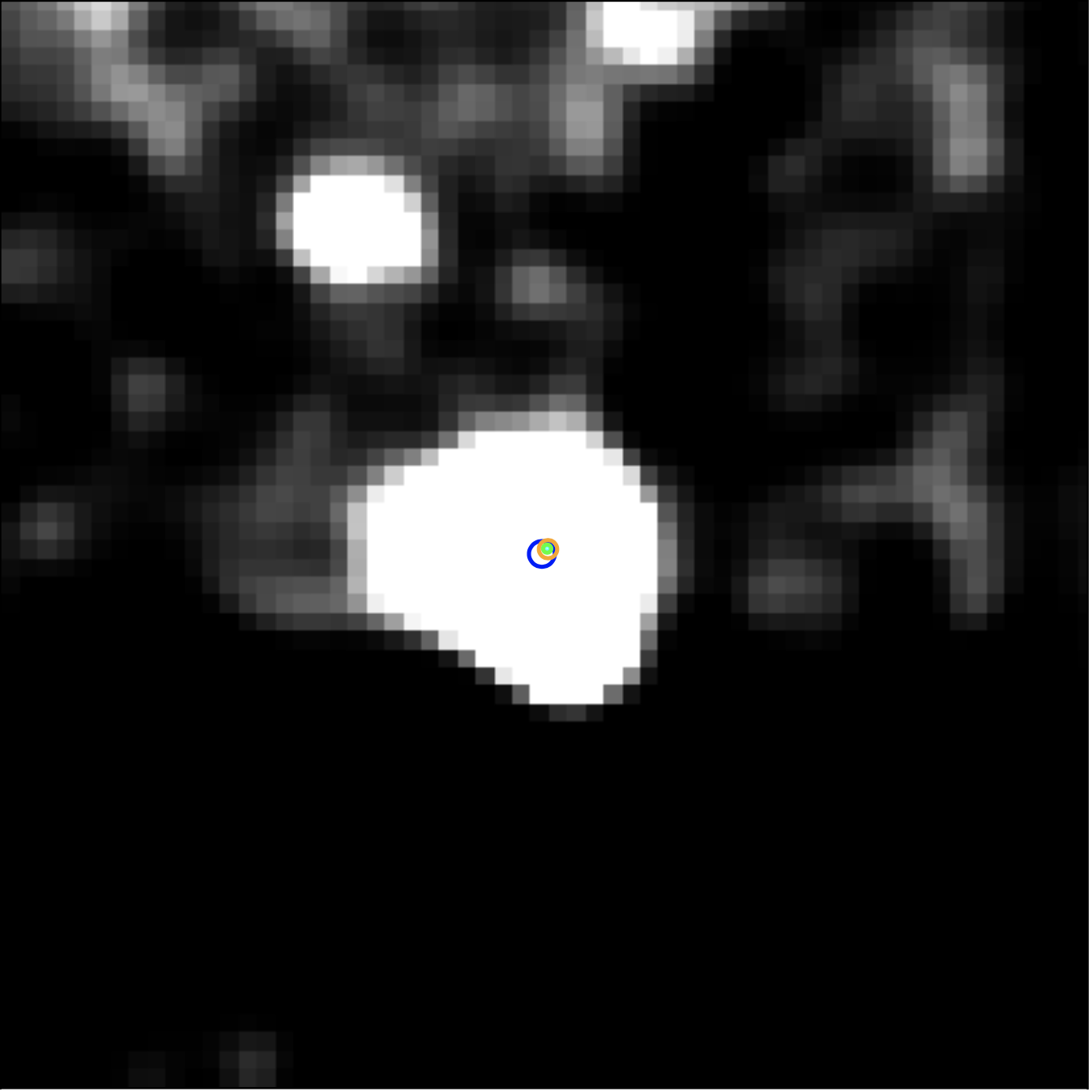}{0.15\textwidth}{XMM00133}\fig{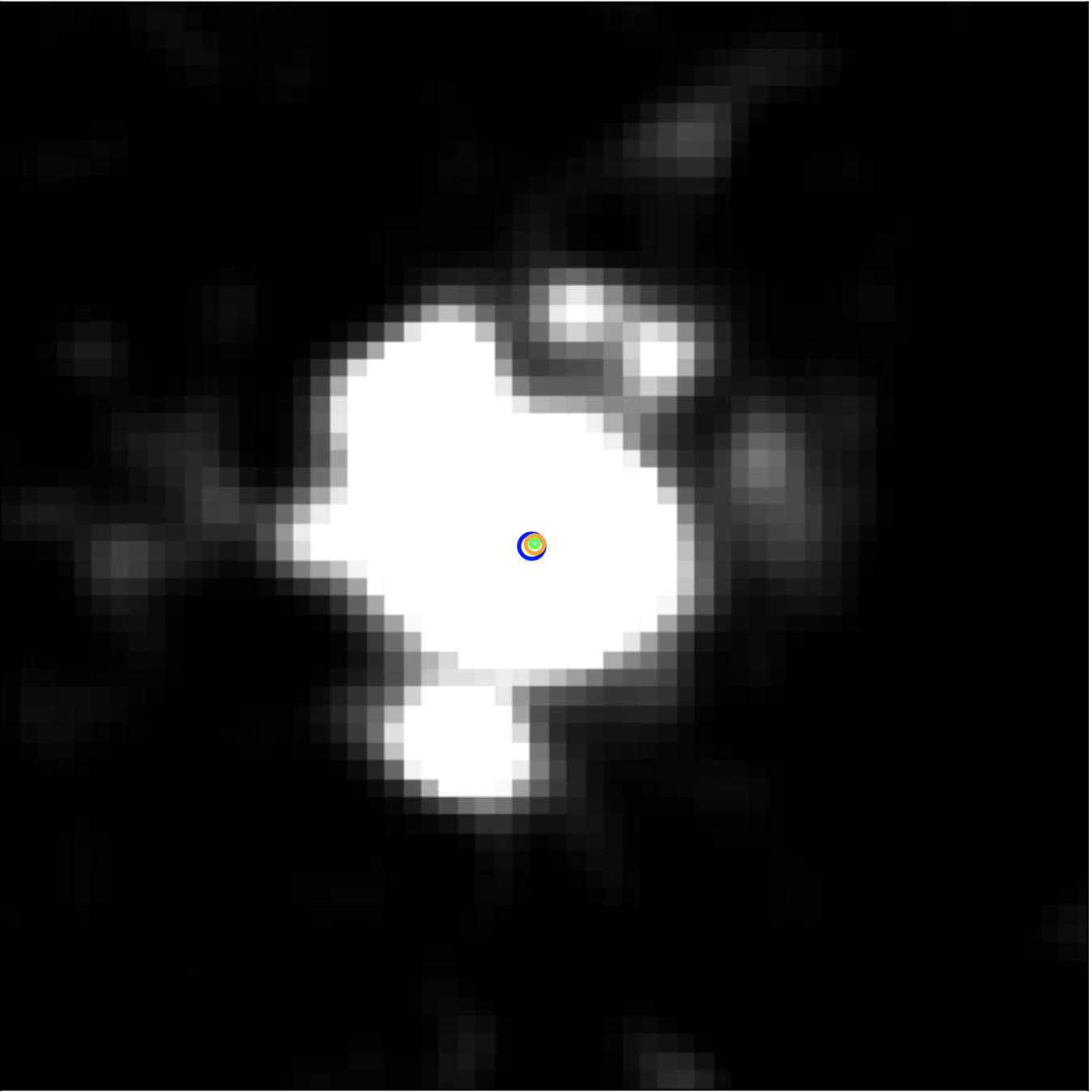}{0.15\textwidth}{XMM00226}\fig{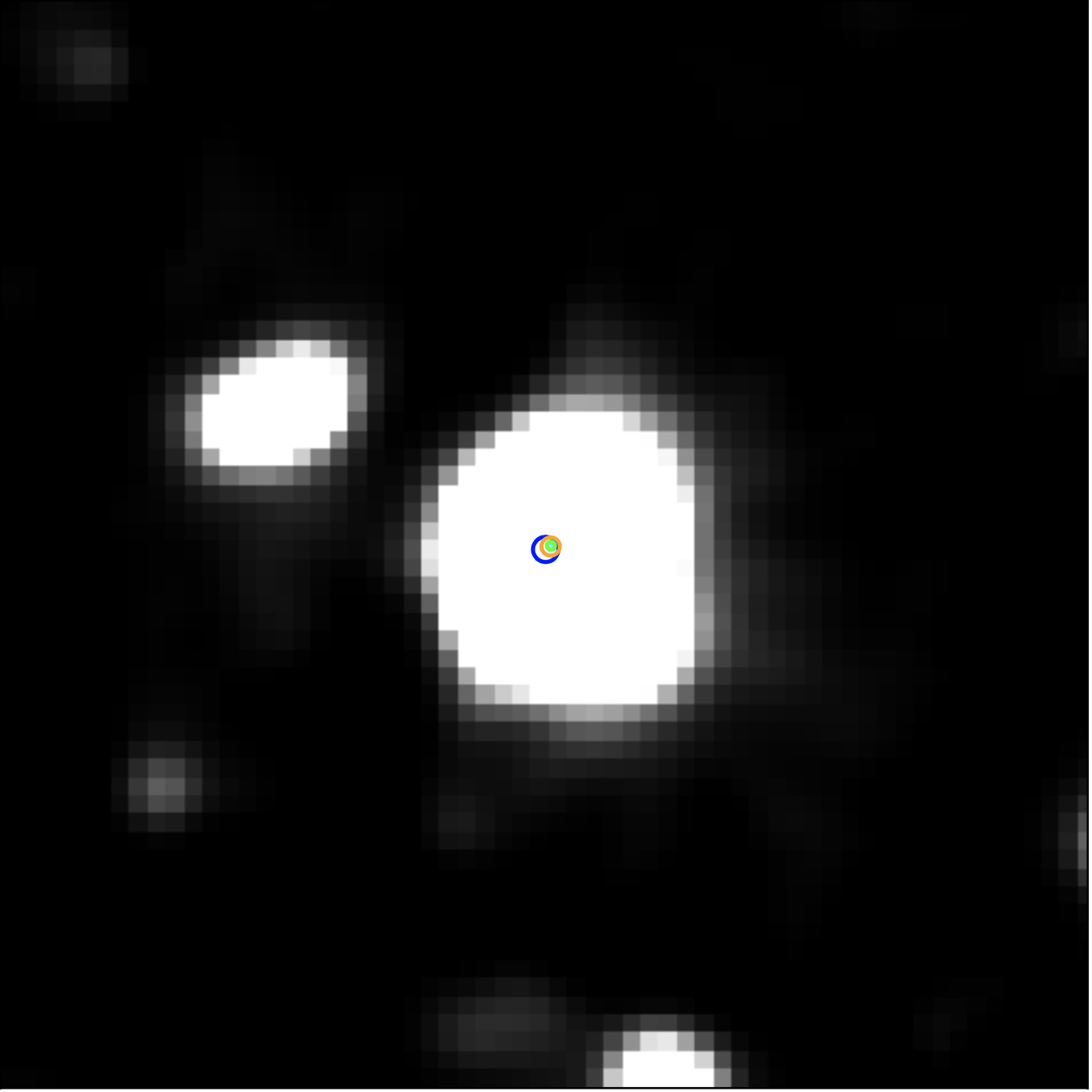}{0.15\textwidth}{XMM00653}}
\gridline{\fig{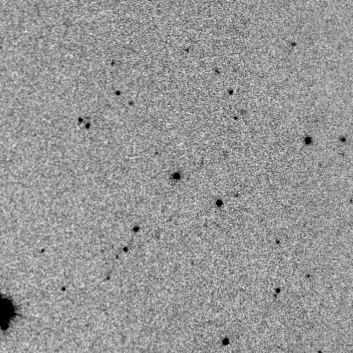}{0.15\textwidth}{}\fig{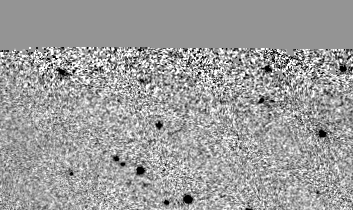}{0.15\textwidth}{}\fig{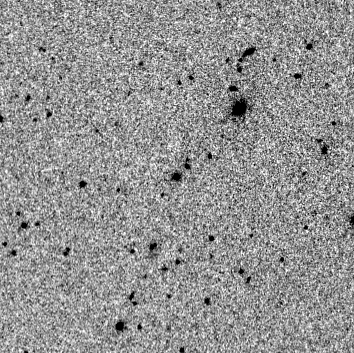}{0.15\textwidth}{}}
\vspace{-0.9 cm}
\gridline{\fig{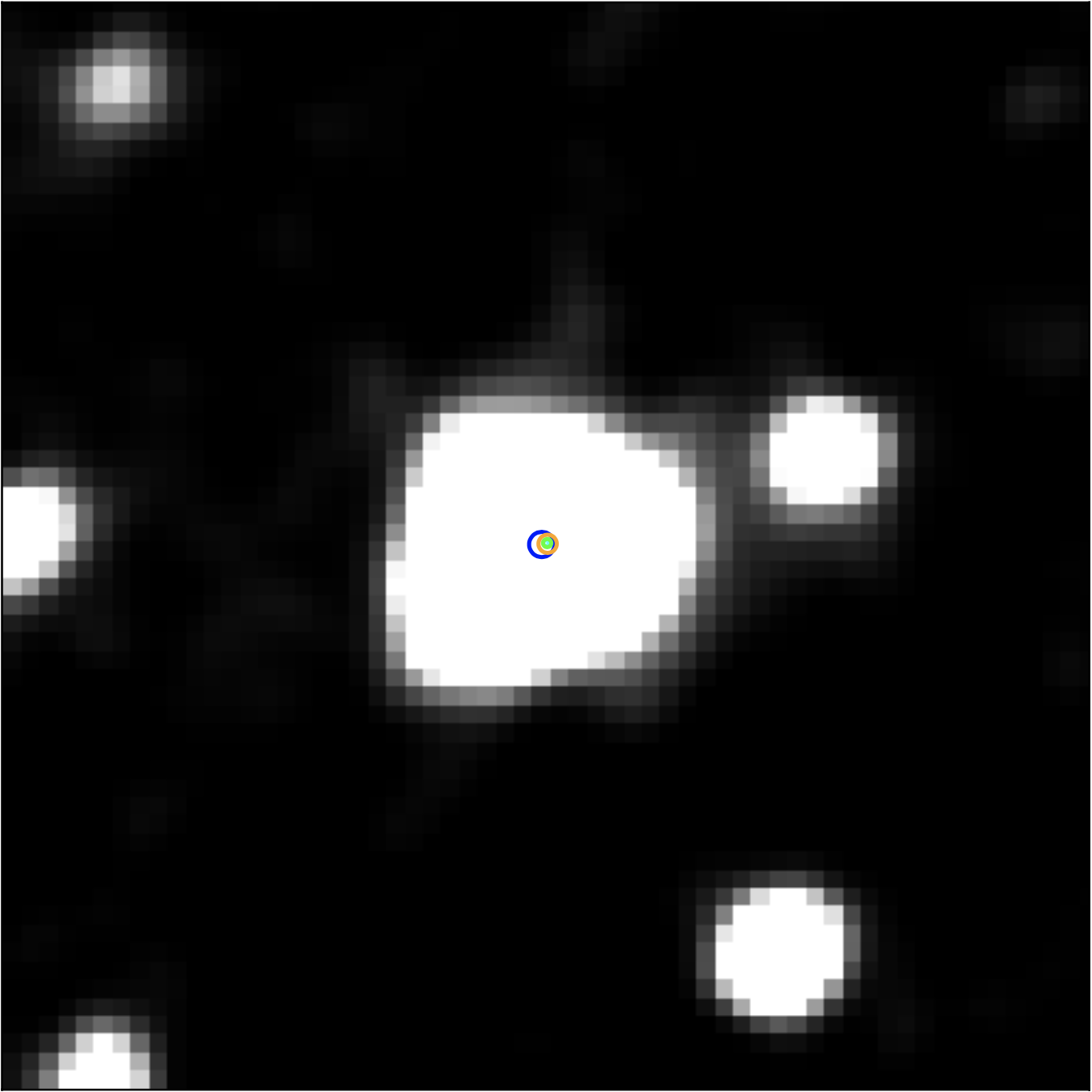}{0.15\textwidth}{XMM00724}\fig{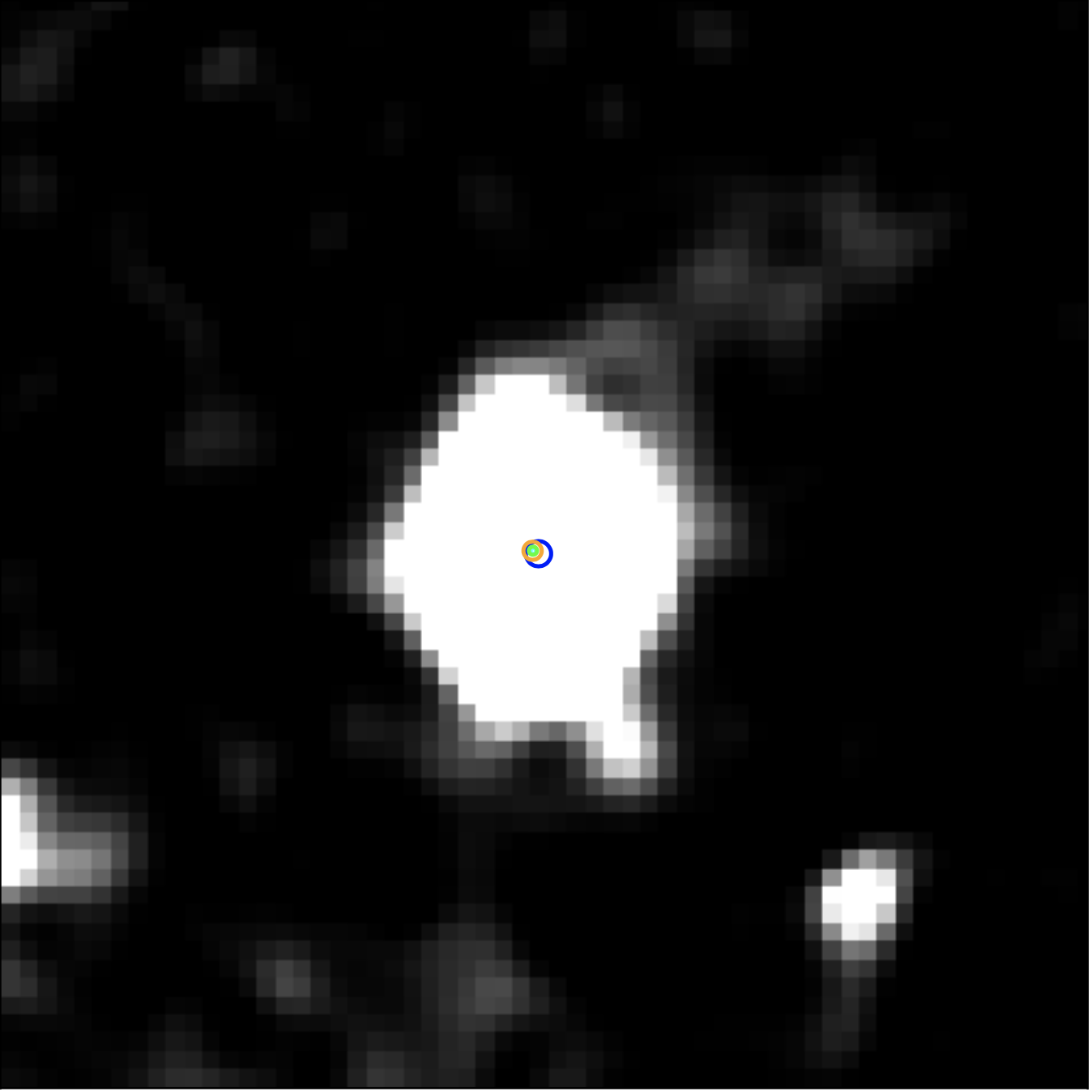}{0.15\textwidth}{XMM01439}\fig{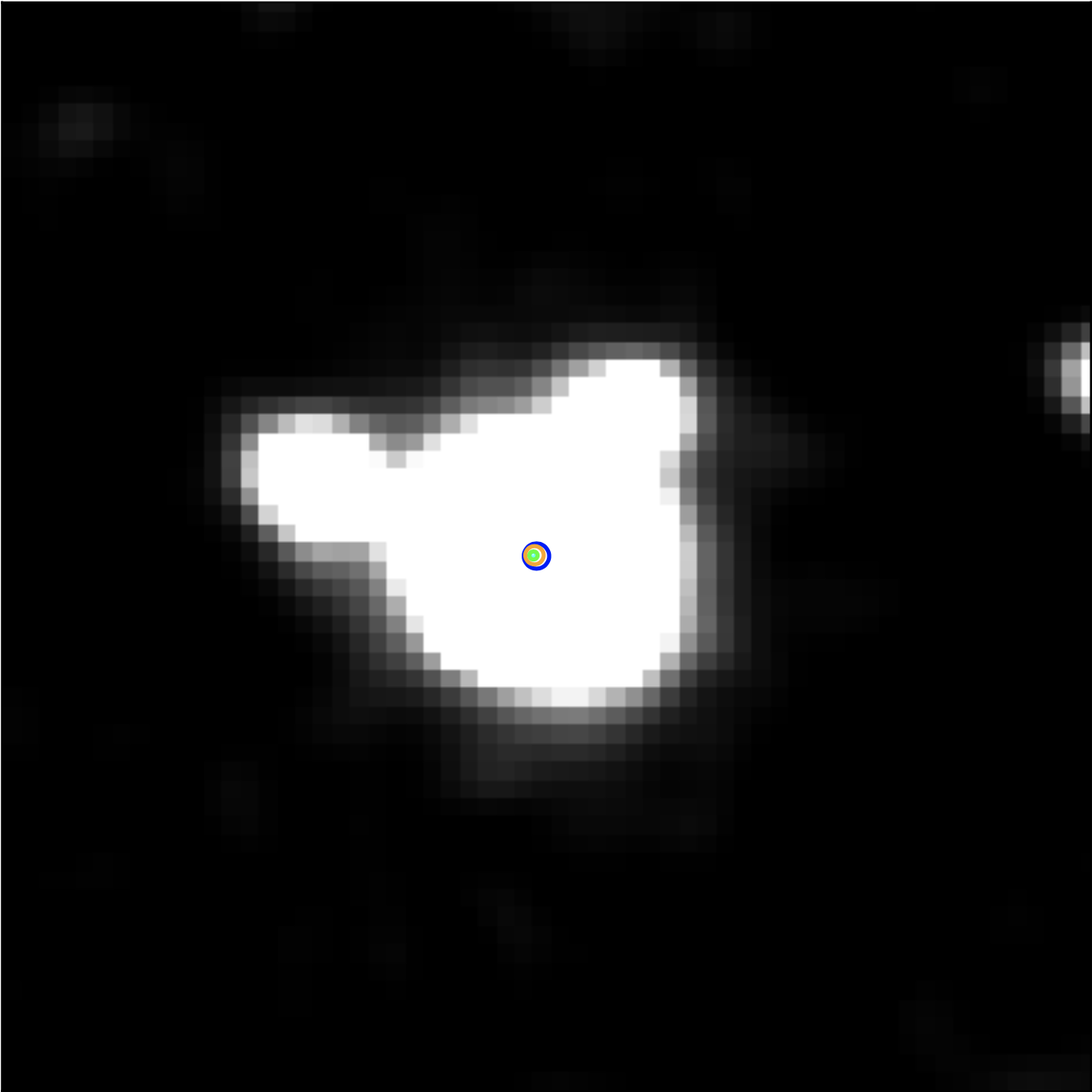}{0.15\textwidth}{XMM02749}}
\gridline{\fig{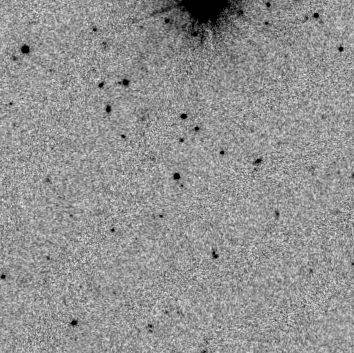}{0.15\textwidth}{}\fig{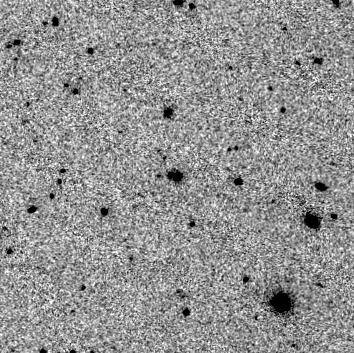}{0.15\textwidth}{}\fig{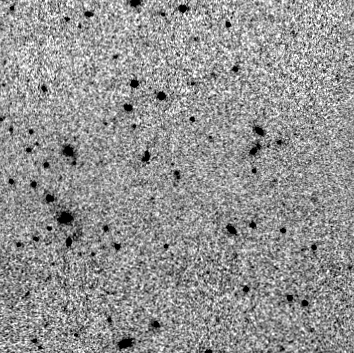}{0.15\textwidth}{}}
\vspace{-0.9 cm}
\gridline{\fig{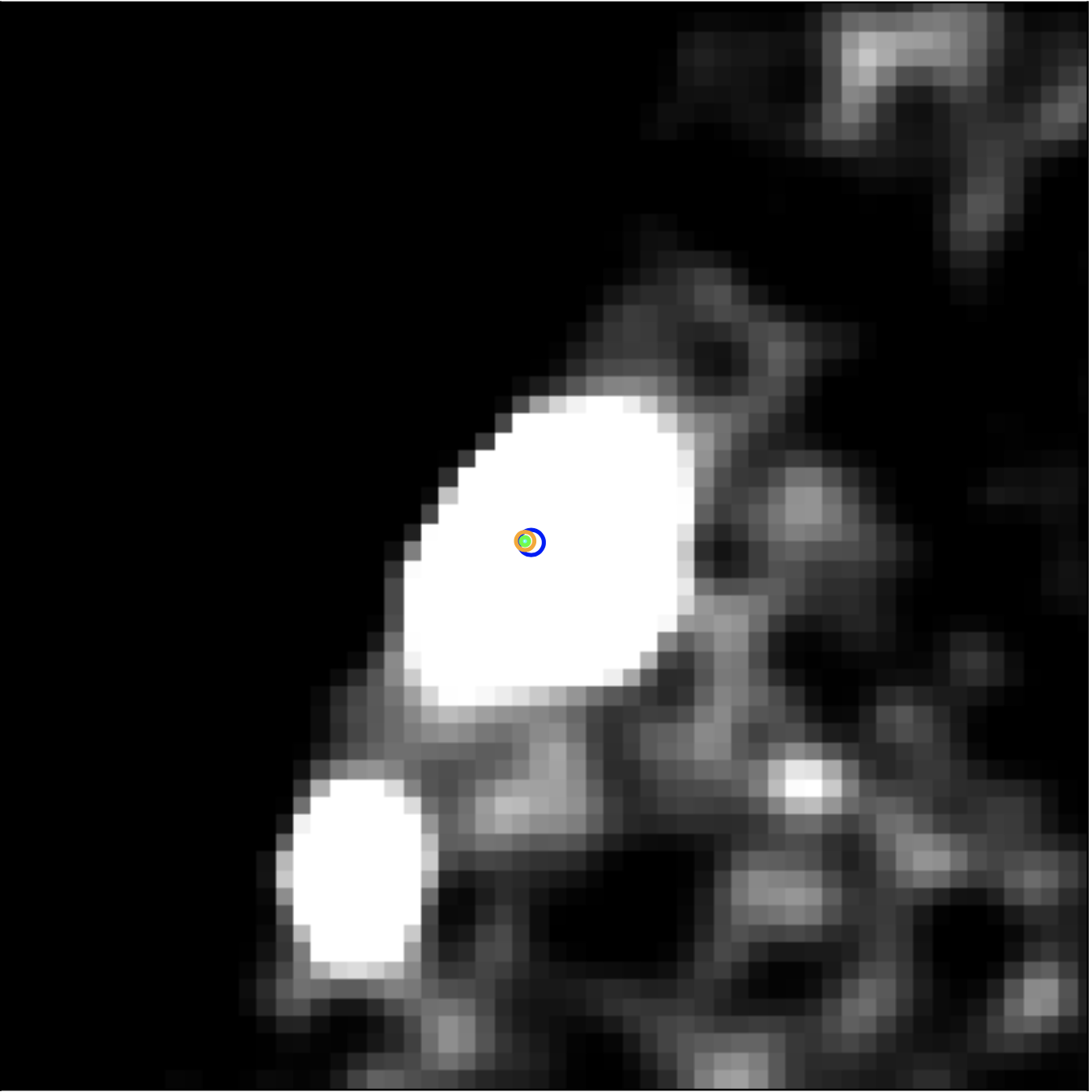}{0.15\textwidth}{XMM03739}\fig{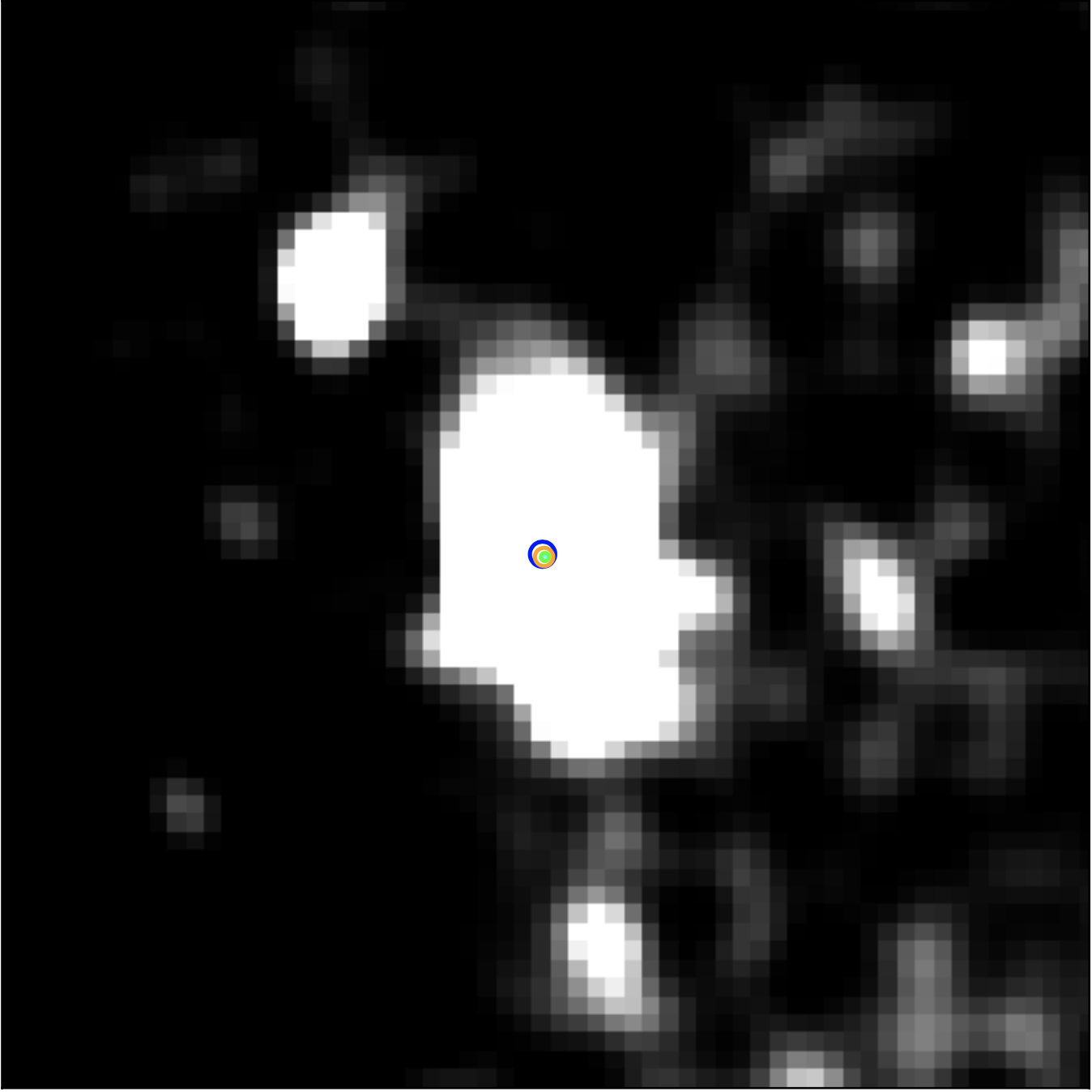}{0.15\textwidth}{XMM05158}\fig{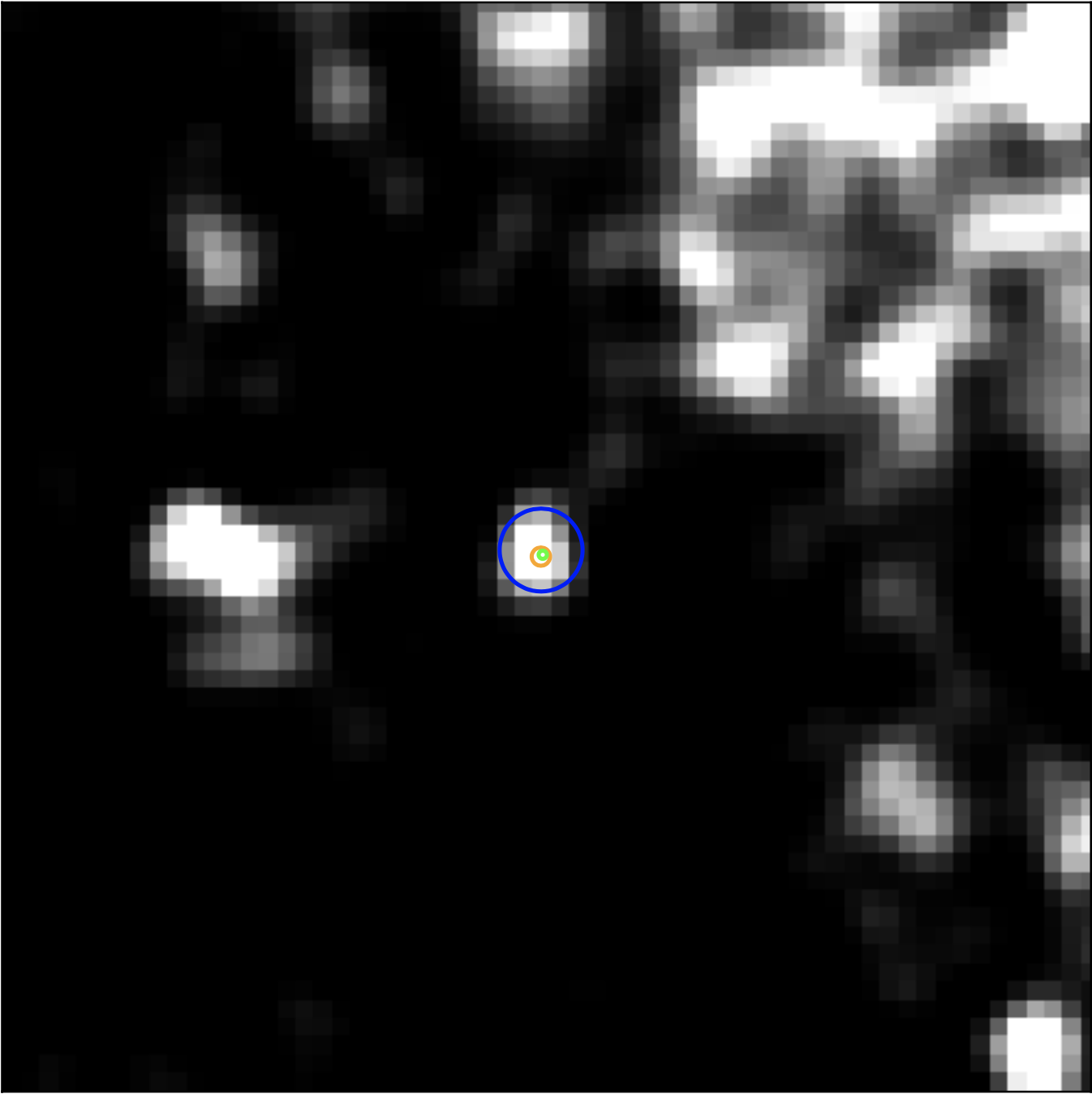}{0.15\textwidth}{XMM05226}}
\caption{Postage stamps of VIDEO K-band sources (top of each row) and the corresponding XMM-SERVS X-ray sources (bottom of each row) that contribute  to the uptick in the XLF at low redshifts and high luminosities. Each image is $2'\times 2'$.}
\label{fig:postagestamps}
\end{figure}

The XLFs in Fig.~\ref{fig:xray_lf_specz} and Fig.~\ref{fig:xray_lf_xmatch} have upticks at high luminosities ($L_X \geq \SI{4e44}{erg\,s^{-1}}$) and low redshifts ($0.63 < z \leq 1.26$).
To investigate this feature  of the XLFs, we identified the  objects--with spectroscopic redshifts since the feature is still present in the `specz only' sample--in the appropriate redshift and luminosity bin.
We looked at these objects' X-ray emission using the postage stamps published in the data artifacts alongside the XMM-SERVS catalog.
Fig.~\ref{fig:postagestamps} include these postage stamps as well as K-band postage stamps from VIDEO.
The \texttt{EMLDETECT} algorithm\footnote{\url{http://xmm-tools.cosmos.esa.int/external/sas/current/doc/emldetect/node3.html}} was used to identify extended X-ray sources.
Many of the objects we investigated appear to be extended X-ray sources that were not correctly categorized by the \texttt{EMLDETECT} algorithm.
The location of this feature at low redshifts echoes the findings of \cite{Clerc2014TheXS} that most of the bright X-ray clusters in the XMM-LSS survey are located at $z < 1.06$.

The redshift-luminosity bins that these objects fall into only have between one and three sources so the misidentification of one X-ray cluster as an AGN dramatically affects the value of the LF.
The uptick of the XLF is also present in the XLF constructed using IR-selected AGN from the comparative sample.
These sources are likely AGN situated in X-ray clusters, leading to an increased X-ray flux and the overestimate of the XLF in these luminosity bins.
We leave these sources in our sample (along with any unidentified extended sources) and attribute the uptick in the LF to them. 

\subsection{Number density}
\begin{figure}[hptb]
\centering
\includegraphics[width=\columnwidth]{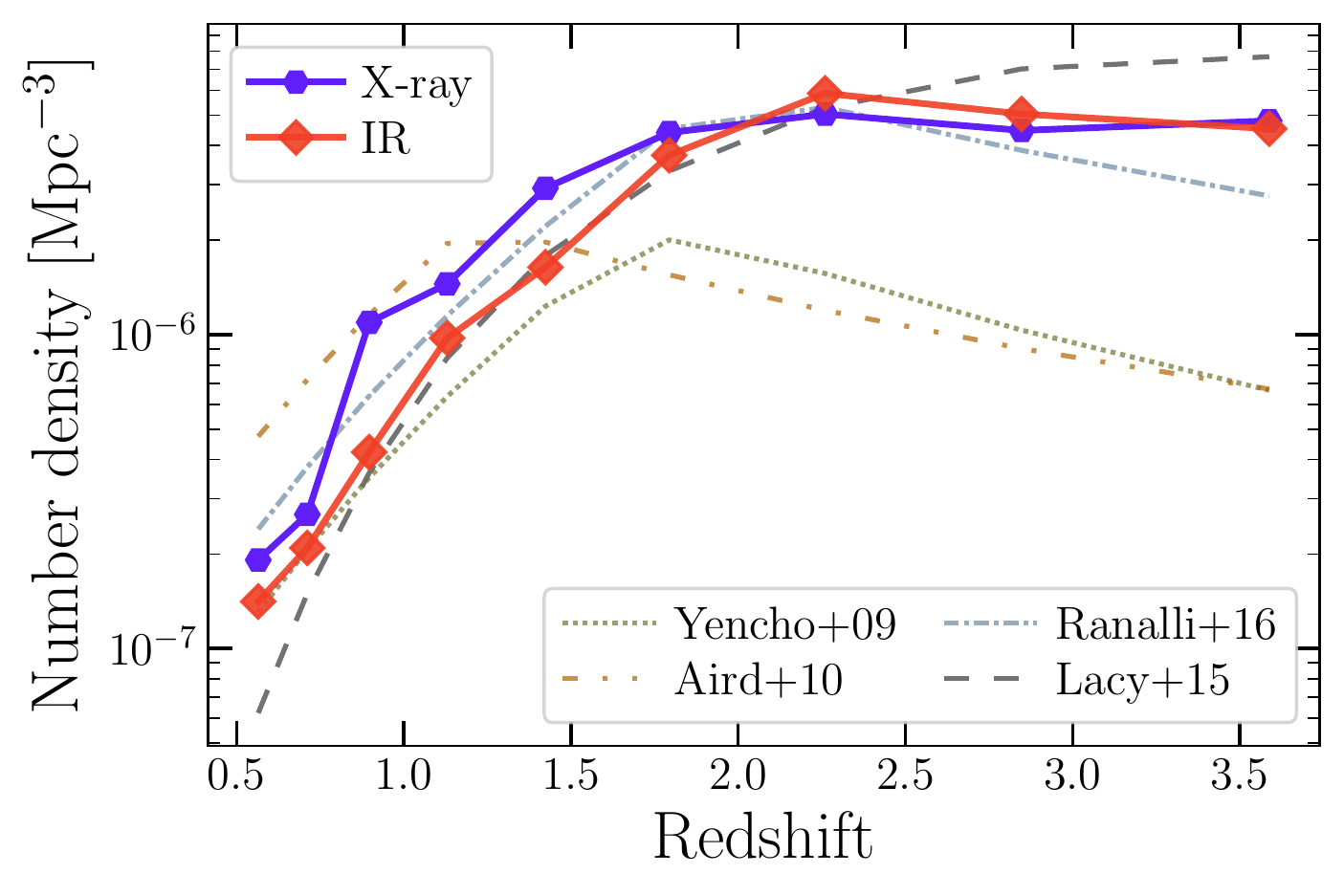}
\caption{Number density of AGN in the luminosity interval $44.5 < \log_{10} L_X < 46$. The number density computed in both the X-ray and IR bands is shown, but note that the luminosity interval for the IR is adjusted using the relationship in Fig.~\ref{fig:lvsl}. Explicitly, the IR luminosity interval is $45.2 < \log_{10} L_{5.8} < 48.1$. In each redshift bin, the LF is fit to a double-power law using orthogonal distance regression which is used to compute the number density for the given luminosity interval. Comparison to previous results from XLFs \citep{yencho_optx_2009, ranalli_210_2016, aird_evolution_2010} and the IRLF from \cite{lacy_spitzer_2015} are shown.}
\label{fig:num_dens}
\end{figure}

The number density tracks the population of AGN over cosmic history.
In Fig.~\ref{fig:num_dens}, we show the number density of AGN computed in both the full X-ray and IRAC3 bands.
We adjust the luminosity in the IR band to correspond to the luminosities in the X-ray by using the empirical relationship from Fig.~\ref{fig:lvsl}.
Additionally, our LF is not supported by the data across the same luminosity interval in all redshift bins.
In order to more accurately compute the integral, we fit the LF in each redshift bin to a double power law of the form
\begin{equation}
\label{eq:dpl}
\frac{d\Phi}{d\log L} = A \left[\left(\frac{L}{L_\star}\right)^{\gamma_1} + \left(\frac{L}{L_\star}\right)^{\gamma_2} \right]^{-1}
\end{equation}
using the orthogonal distance regression routine from \texttt{scipy.odr}.
This fit is not meant to robustly determine a parametric estimate of the LF nor its evolution as a function of redshift; we are using this parametrisation to simply reproduce the shape of the LF.
Because of this limitation, we choose a luminosity interval $44.5 < \log_{10} L_X / ({\rm erg s^{-1}}) < 46$ so as not to rely too heavily on extrapolation of these fits.
Changes of roughly \SI{0.5}{dex} to this luminosity interval did not appreciably impact the results.
Larger changes push the luminosity interval to regions that are not well supported by our data.
We leave a robust parametric estimate and further implications of that LF for future work.

The number densities computed from the IRLF and XLF shown in Fig.~\ref{fig:num_dens} are in good agreement.
These number densities peak around $z\sim 2.25$ and are most consistent with the results of \cite{ranalli_210_2016}.
We provide comparisons to previous results \citep{aird_evolution_2010, yencho_optx_2009, silverman_luminosity_2008, brandt_xmm-newton_2009} in the discussion.

\subsection{Luminosity density}
We also compute the bolometric luminosity density of the AGN and present a comparison to the literature \citep{aird_evolution_2010, brandt_xmm-newton_2009, lacy_optical_2007}.
Similar to our treatment of the number density, we fit the LF in each redshift bin to the double power law in Eq.~\ref{eq:dpl} to have support over the entire luminosity interval.
To compute the bolometric luminosities, we use simple bolometric corrections from \cite{lacy_spitzer_2015} for the IR.
For the X-ray bolometric corrections, we use the luminosity-dependent bolometric corrections from \cite{duras_universal_2020}.
The bolometric luminosities are modified by a multiplicative factor $L_\mathrm{bolo} = C_\nu\times L_\nu$ where $C_{5.8} = 10$ and
\begin{equation}
C_X(L_X) = a\left[1+\left(\frac{\log_{10}{L_X/L_\odot}}{b}\right)^c\right]
\end{equation}
where $a=15.33$, $b=11.48$, $c=16.2$.

\begin{figure*}[hptb]
\centering
\includegraphics[width=0.8\textwidth]{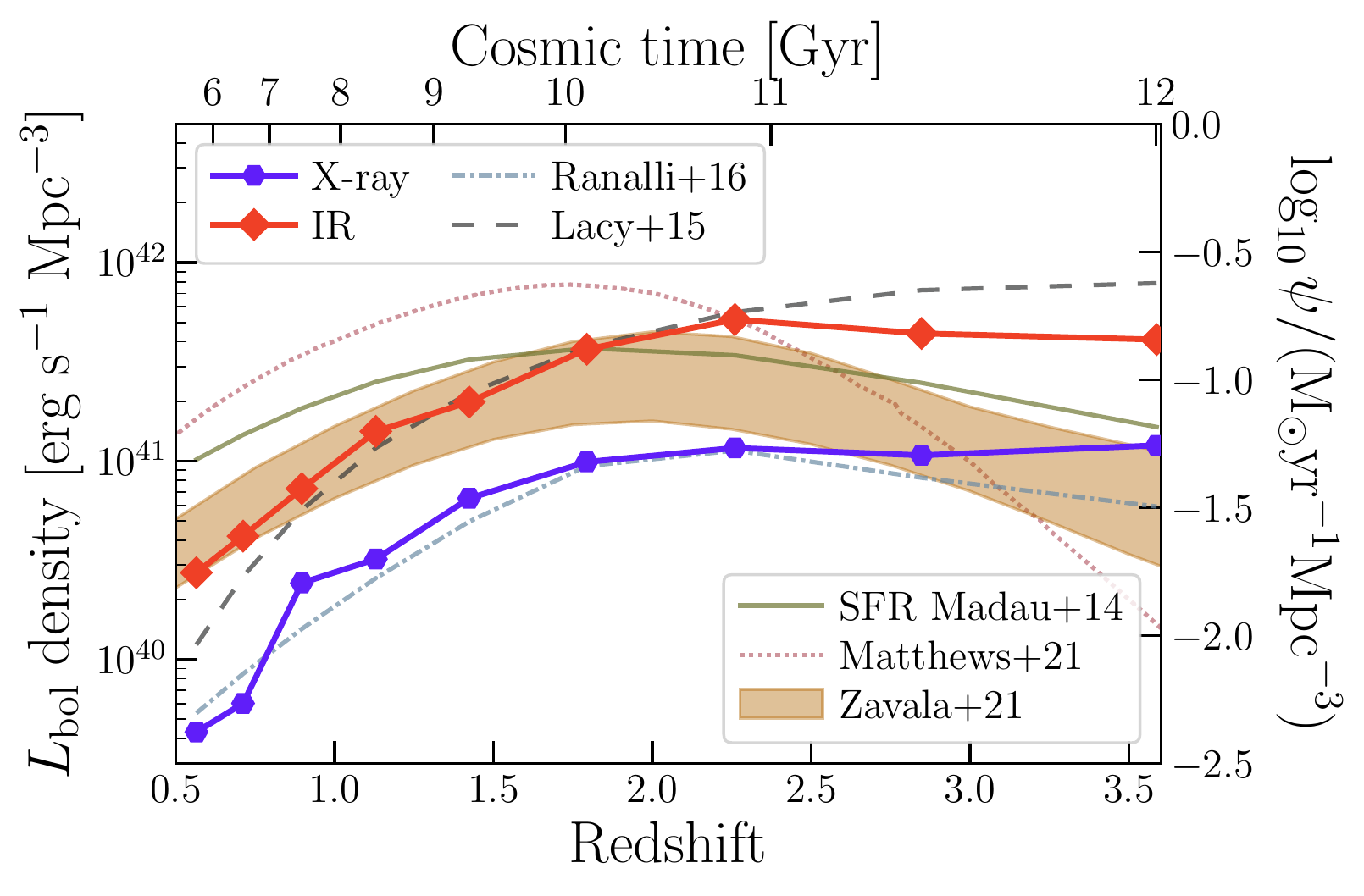}
\caption{Bolometric luminosity density computed in the X-ray and IR for the bolometric luminosity interval $45.8 < \log_{10}L_\mathrm{bolo} < 47.8$ (this is the corresponding interval of the bolometric luminosity to that in Fig.~\ref{fig:num_dens}). A simple bolometric correction of $C_{5.8}=10$ is taken from \cite{lacy_spitzer_2015}. Luminosity dependent X-ray bolometric corrections are taken from \cite{duras_universal_2020}. As with the number density, we fit the LF in each redshift bin to a double power law. We include comparisons of the luminosity density from previous results \citep{ranalli_210_2016,lacy_optical_2007} as shown in the upper left legend.  Additionally shown on the right side y-axis (and the lower right legend) are the star formation rate densities (SFRD) from \cite{madau_cosmic_2014}, \cite{zavala_evolution_2021}, and the \SI{1.4}{GHz} radio-derived SFRD from \cite{matthews_cosmic_2021}.}
\label{fig:lum_dens}
\end{figure*}

% \begin{figure*}[h!]
%     \centering
%     \includegraphics[width=\textwidth]{figures/combined_lf.pdf}
%     \caption{Bolometric LF for 30 arcsec cross match. This LF includes K-corrections and bolometric corrections for both X-ray and IR data.}
%     \label{fig:lf}
% \end{figure*}

\section{Discussion}
\label{sec:discussion}
\subsection{Comparison of the LFs with previous results}
Our estimates of the AGN LF in both the X-ray and IR bands improve significantly on those of previous results.
The IR sample is a factor of $\sim 6$ larger than the strictly spectroscopically-identified sample of \cite{lacy_spitzer_2015}.
Additionally, our IR sample has sources with luminosity $L_{5.8} < \SI{1e44}{erg~s^{-1}}$ and significantly more AGN across the entire luminosity range at $z\gtrsim 1.5$.
The X-ray sample is a factor of $\sim 2$ larger than the \cite{ranalli_210_2016} sample and has more high-luminosity sources ($L_X \gtrsim \SI{1e45}{erg~s^{-1}}$) at $z>1.5$.
We also include a novel method of accounting for catastrophic failures of photometric redshift estimates and robust comparisons of our LFs using subsamples of the full catalogs.
Both LFs are consistent with previous results.
However, the IRLF does indicate flattening at the faint end at $z\gtrsim2$ and the XLF favors higher number densities at $z\gtrsim1.5$, more consistent with the recent results of \cite{ranalli_210_2016}.

The comparison of our IRLF to the \cite{lacy_spitzer_2015} model shows good agreement, in particular in the higher luminosity regime.
Our sample has better faint end coverage at lower redshift and significantly better coverage across the entire luminosity interval at higher redshifts.
The binned IRLF obtained here is consistent with the extrapolation of their model to intermediate and higher redshifts as well as lower luminosities.
The \cite{lacy_spitzer_2015} sample was insufficient to constrain the IRLF at low luminosities ($\lesssim\SI{1e45}{erg~s}$) for $z\gtrsim 1.0$. 
However, our IRLF shows that their power-law evolution model is reliable across a larger redshift and luminosity range.
For redshifts $z<1.0$ and luminosities $L_{5.8}<\SI{3e44}{erg~s^{-1}}$, our IRLF overestimates the \cite{lacy_spitzer_2015} model.
The \cite{lacy_spitzer_2015} had strict criteria for AGN identification.
Their spectroscopic sample was potentially incomplete because emission line ratios of low luminosity objects are hard to get.
This discrepancy at low luminosity and low redshifts may be due to the inclusion of faint AGN in our sample but this region is also near the flux limits of SWIRE (thus where incompleteness is highest) so the error bars are likely underestimated.
Further work is needed to confirm the low luminosity IRLF.
Additionally, there is disagreement at low luminosities ($L_{5.8}\lesssim\SI{1e45}{erg~s^{-1}}$) up to redshift $z<2.5$ between the full sample and the specz only sample.
Objects with low luminosities have low emission line strength so these objects are more likely to be left out of the specz only sample.
The discrepancy between the X-ray selected IRLF and the full sample IRLF at $L_{5.8}\lesssim\SI{1e45}{erg~s^{-1}}$ and $z<2.0$ may be due to obscuration (IR sources are lacking X-ray counterparts or they are too faint to be selected) or to intrinsically-lower X-ray luminosities due to these sources being in high-accretion modes. 

The IRLFs in Fig.~\ref{fig:ir_lf_specz} indicate a flatter faint end slope of the LF at roughly $z>2$.
The significance of this flattening increases as redshift increases.
Comparing the lowest luminosity point from our LF to literature values, we estimate the significance of this flattening.
At $z\sim 2.75$, the flattening is significant at $\sim 2 \sigma$.
This discrepancy at higher redshifts may be due to insufficient sampling in the \cite{lacy_spitzer_2015} sample or a problem with their model.
The model did not allow freedom for the faint end slope to vary in their fits so any flattening of the faint end at high redshifts would not be apparent from their results.
This flattening is consistent with LF results in optical bands \citep{ross_sdss-iii_2013}.
The flattening suggests that the number density of fainter AGN remains roughly constant across these redshifts while the number density of very luminous AGN decreases as redshift decreases.
However, it is possible that the flattening of the faint end slope of the IRLF is due to losing lower luminosity AGN that are outshone by their host galaxies.

The XLF has been well studied.
The close agreement of our LF to previous results is encouraging.
Our incompleteness corrections and treatment of photometric redshift uncertainties are able to produce consistent results with \cite{ranalli_210_2016, aird_evolution_2010}.
We note that at higher redshifts, our XLF favors higher values outside of the confidence intervals of \cite{aird_evolution_2010} but consistent with those of \cite{ranalli_210_2016}.
Agreement suggests that the LADE model from \cite{ranalli_210_2016} accurately models the AGN population from $0<z<4$ and with luminosities $43.3 < \log_{10}L_{X} \lesssim 46$.

\subsection{Implications of the number \& luminosity densities}
In Fig.~\ref{fig:num_dens}, we show our number density and comparisons to previous studies.
We find good agreement in the number density computed using the XLF and IRLF, but we find some discrepancies with previous work.
In particular, our results favor higher number densities as found in \cite[and]{lacy_spitzer_2015, ranalli_210_2016} at $z\gtrsim1.75$ than most previous work \citep{yencho_optx_2009, silverman_luminosity_2008, aird_evolution_2010}.
However, many previous studies lack the depth and statistics to constrain the LF at high redshift.
Specifically, strictly spectroscopic samples may be chosen due to the high accuracy of their redshifts, but, as emission line luminosity is correlated with AGN accretion rate, highly complete spectroscopic AGN samples tend to be hard to obtain at low AGN luminosities, especially at high redshifts where host galaxy features are difficult to measure.
For smaller samples, the LF in low redshift bins is constrained by a parametric model governing the evolution of the LF.
Studies with few AGN at high redshift will rely too heavily on the extrapolation of the low redshift LF and its evolution rather than a constraint based on the data.
At redshift $z\gtrsim2.5$, our number densities are constrained by data (our number density from the IRLF does rely on extrapolation to the high end of the luminosity interval, but this region will contribute negligibly to the number density since the LF is decreasing rapidly) whereas \cite{lacy_spitzer_2015} relies on extrapolation of the model across the entire luminosity interval and \cite{ranalli_210_2016} extrapolates for $L_{X}\gtrsim\SI{1e45}{erg~s}$.
We confirm their results at low redshifts and extend the estimates to $z\sim4$.

The AGN luminosity density derived from the XLF is in good agreement with \cite{ranalli_210_2016} up to $z\sim 2.25$ (and likewise for the luminosity density derived from the IRLF and \cite{lacy_spitzer_2015}).\footnote{We found that the AGN luminosity density estimates based on our XLF were larger than those found in \cite{aird_evolution_2010} and \cite{yencho_optx_2009} which only had spectroscopic samples.}
However, we note a factor of $\sim3$ difference between the luminosity densities derived using the XLF and IRLF. 
We do not correct for obscuration so the unobscured X-ray luminosity density is likely higher than presented here.
However, \cite{ranalli_210_2016} include corrections for absorption which suggests that these corrections are not a large factor.
These absorption corrections are by their nature unable to correct for objects missing from the sample.
So the discrepancy between the luminosity densities is likely due to the simple bolometric corrections used for the IR or due to heavily obscured AGN being missed in X-ray surveys.
In Fig.~\ref{fig:lum_dens}, the bolometric luminosity density computed using the XLF peaks at $z\sim 2.25$ and flattens off, falling slightly at higher redshifts. 
The luminosity density computed using the IRLF peaks around the same redshift.

\subsection{Comparing the AGN luminosity density with the SFRD}
For comparison, we also plot the cosmic star formation rate density (SFRD) from \cite{madau_cosmic_2014}, \cite{zavala_evolution_2021} and the radio-derived SFRD from \cite{matthews_cosmic_2021} in Fig.~\ref{fig:lum_dens}.
This peaks at $z\sim 1.75$, roughly $2$~Gyr later than the luminosity density at $z\sim 2.25$.
From redshifts $0 < z \lesssim 1.75$, AGN luminosity density and SFRD are both decreasing.
Assuming that accretion luminosity is linearly proportional to the accretion rate, the black hole mass growth rate falls by roughly an order of magnitude over the same time frame. 
The SFRD only falls by roughly half an order of magnitude over the same time frame.
The co-moving mass density of molecular gas is roughly flat between $3 > z > 2$ but decreases by roughly \SI{0.7}{dex} between $2>z>0.5$ \citep{yan_alpine-alma_2020, garratt_cosmic_2021}.
This is consistent with the idea that both AGN and SF are governed by the availability of free gas.
Fundamentally, this is due to collisional material (gas) being converted into collisionless material (stars).
% Since gas needs to collapse to scales of $\sim\SI{100}{pc}$ in the case of SF and $\sim\SI{0.1}{pc}$ scales in the case of AGN, this suggests that processes that collapse gas to sub-parsec (i.e. AGN) scales become relatively less efficient than those necessary for star formation. 
Observations of cosmic downsizing \citep{2005ApJ...619L.135J} suggest that the bulk of star formation moves from the massive galaxies that host the massive black holes whose activity dominates the AGN luminosity density to smaller galaxies with smaller black holes, or perhaps none at all.
The uncertainties in the peaks of the luminosity densities are large enough such that we cannot claim this robustly.

The AGN luminosity density is constant from $3.5 < z < 2$ as has been found in previous studies \citep{lacy_spitzer_2015, ranalli_210_2016}.
In the same time frame, the SFRD grows by 0.5-1.5 dex.
We note that the AGN luminosity density and SFRD are uncertain at these redshifts, but if this trend is confirmed it would suggest an independent evolutionary pathway for early AGN and star formation.
While this difference in behavior at higher redshifts could be due to early and extremely efficient creation of SMBHs through a different channel than at $z\sim 2$, we think is unlikely.
The most plausible explanation is that once host galaxies become sufficiently large, growth of SMBHs and star formation are significantly quenched \citep{peng_mass_2010}.
AGN in more massive galaxies can accrete, but at a lower rate relative to the Eddington limit.
Additionally, our data suggest that black hole mass growth precedes host galaxy growth at higher redshifts.
This means that at $z \gtrsim 3$ we expect a population of galaxies with black hole masses larger than the local $M-\sigma$ relationship suggests \citep{magorrian_demography_1998, kormendy_coevolution_2013, merritt_mbh-sigma_2001}.
This is consistent with observations of high-redshift quasars ($z\gtrsim 6$) \cite[eg,]{2018ApJ...854...97D, neeleman_kinematics_2021} that have larger black hole-host galaxy ratios than the local population of galaxies.

\section{Summary \& Conclusions}
\label{sec:conc}
We present binned estimates of the IRLF in the MIR IRAC3 band and XLF in the full band in the XMM-LSS field using HELP and XMM-SERVS data.
There are 5071 AGN used to construct the XLF and 5918 AGN used to construct the IRLF in the redshift range $0.5 \leq z \leq 4.0$ and the luminosity range \SIrange[range-phrase=--, range-units=single]{2e43}{3e46}{erg~s^{-1}}.
Using a subsample of our sources in a comparative field, we use X-ray selection criteria for the IRLF and vice versa to provide a crosscheck on our analysis.

The IR sample is larger by a factor of $\sim 6$ than recent studies (eg., \cite{lacy_spitzer_2015}) with better coverage at lower luminosities ($L_{5.8}\leq\SI{1e44}{erg~s^{-1}}$ at $z<1.5$ and $L_{5.8}\leq\SI{3e45}{erg~\s^{-1}}$ at $z\geq1.5$).
The X-ray sample is larger by a factor of $\sim 2$ than more recent studies (eg., \cite{ranalli_210_2016}) and a factor of $\sim 4$ for others (eg., \cite{silverman_luminosity_2008}) with better coverage of high luminosities ($L_X\geq\SI{1e45}{erg~s^{-1}}$) at $z\gtrsim1.5$.
The sample also has more AGN at high redshifts ($z\geq 2.5$), allowing us to extend the estimate of our LFs to higher redshifts than many previous results \citep{aird_evolution_2010, yencho_optx_2009}.

Our analysis includes a novel method to account for photometric redshift uncertainties.
A MCMC chain is constructed using redshifts pulled from PDFs that are generated from a subsample of AGN.
These PDFs are often multimodal with long tails, accounting for catastrophic failure of photometric redshift fitting routines.
This method is accompanied by crosschecks of the LFs.
We construct the LF in a subfield using a comparative sample. 
Then X-ray selection criteria can be used to select a sample for the IRLF and vice versa.
We also construct the LFs using a strictly spectroscopic sample and construct the LFs with Gaussian uncertainties to ensure we handled photometric redshift uncertainties properly without introducing a bias.
Generally, we find that our LFs are consistent across these crosschecks and with previous studies.
We draw the following new conclusions:
\begin{itemize}
\item The IRLF has the standard double power law shape. At low redshift, the knee is nearly indiscernible. For the first time, our data allows us to probe the faint end slope. At $z\sim 2$, we find evidence of flattening in the faint end slope. This flattening is $\geq 3\sigma$ in the last three redshift bins.
\item The XLF has a more pronounced knee that evolves to higher luminosity with increasing redshift. Our data does not probe the faint end slope at higher redshifts. However, the bright end XLF is in excess of previous studies for $z\gtrsim 2.5$. The precision of our results is higher than previous studies using photometric redshifts (eg., \cite{ranalli_210_2016}).
\item We construct the LF in a subfield using a comparative sample. We provide a comparison in both bands of the LF using X-ray and IR selection criteria. The comparative LF agrees with the full sample LF, but we see a small excess in the faint end of the LF when using X-ray selection criteria in both the XLF and IRLF.
\item The number density grows until $z\sim 2$ and then begins to fall slightly. We find reasonable agreement between the number density computed with IRLF and XLF and best agreement with the recent results of \cite{ranalli_210_2016}. However, our results refine their estimates because our LFs have consistently smaller uncertainties.
\item The bolometric luminosity density peaks at $z\sim 2.25$ before the SFRD at $z\sim1.75$. The bolometric luminosity density inferred from the IRLF is a factor of $\sim3$ larger than that inferred from the XLF at all redshifts. 
\item The constancy of the luminosity density at $z\gtrsim2$ implies that quenching of AGN activity occurs once host galaxies becomes sufficiently large.
\end{itemize}

We leave further understanding of the evolution of the LF and more direct comparisons between the XLF and IRLF to future work.

\section*{acknowledgements}
The National Radio Astronomy Observatory is a facility of the National Science Foundation operated under cooperative agreement by Associated Universities, Inc.
Basic research in radio astronomy at the U.S. Naval Research Laboratory is supported by 6.1 Base Funding.
JR and DF are supported by NSF grant AST-1934744.

% \begin{acknowledgments}
% JR is supported by NSF grant AAAAAAA.
% \end{acknowledgments}

% \software{Astropy \citep{http://dx.doi.org/10.1051/0004-6361/201322068},
%           Matplotlib \citep{https://doi.org/10.1109/MCSE.2007.55},
%           NumPy \citep{https://doi.org/10.1038/s41586-020-2649-2},
%           SciPy \citep{https://doi.org/10.1038/S41592-019-0686-2}
%           }}
\bibliography{references,more_refs}

\appendix
\section{Using Gaussian uncertainties for photometric redshifts}
\label{app:gauss}
In \S\ref{sec:mc} we present a novel method to account for photometric redshift uncertainties.
We attempt to encode the smaller uncertainties of any template-fitting redshift routine due to the inherent limitations of matching many real astrophysical sources to a limited number of templates as well as the larger uncertainties due to the occasional catastrophic failures of these fitting routines.
However, to get an empirical estimate of these uncertainties, we restrict ourselves to the sample of sources with both spectroscopic and photometric redshifts.
This analysis necessarily assumes that this subsample of sources is representative of the full sample.
Of course, this assumption cannot be completely true: objects with spectroscopic redshifts to be bright and strong-lined.
So this assumption may introduce some biases.
Our task then is to understand the effect of this bias on the LF.

To do so, we rerun our analysis and construct an IRLF and XLF for the full samples using the same corrections.
Rather than our PDF treatment, we use Gaussian uncertainties for each source assuming the relationship $\sigma(z) = 0.2 \times (z + 1)$.

\begin{figure}[h]
\centering
\includegraphics[width=\textwidth]{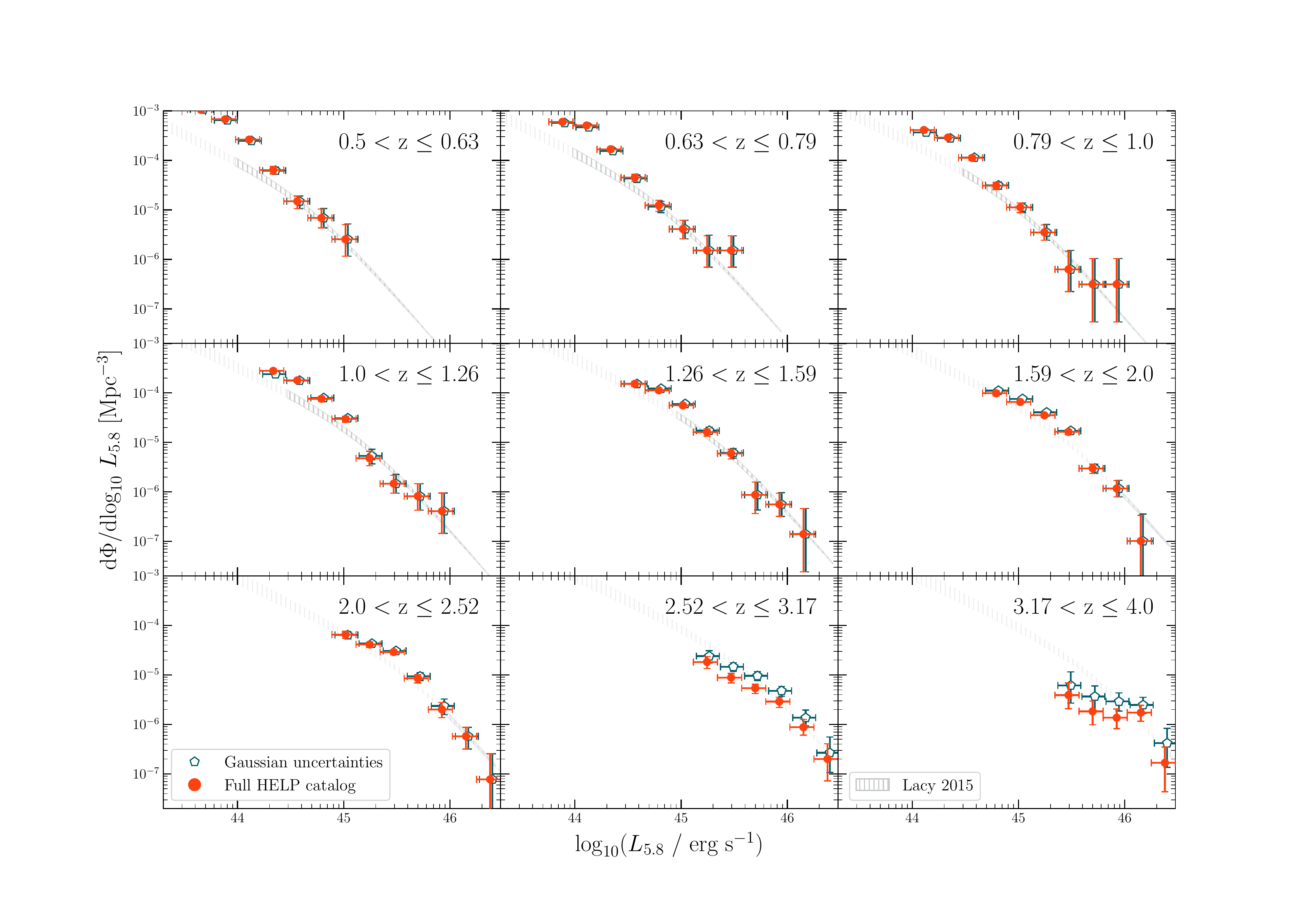}
\caption{IRLF with Gaussian uncertainties on the photometric redshifts.}
\label{fig:lf_ir_gauss}
\end{figure}

\begin{figure}[h]
\centering
\includegraphics[width=\textwidth]{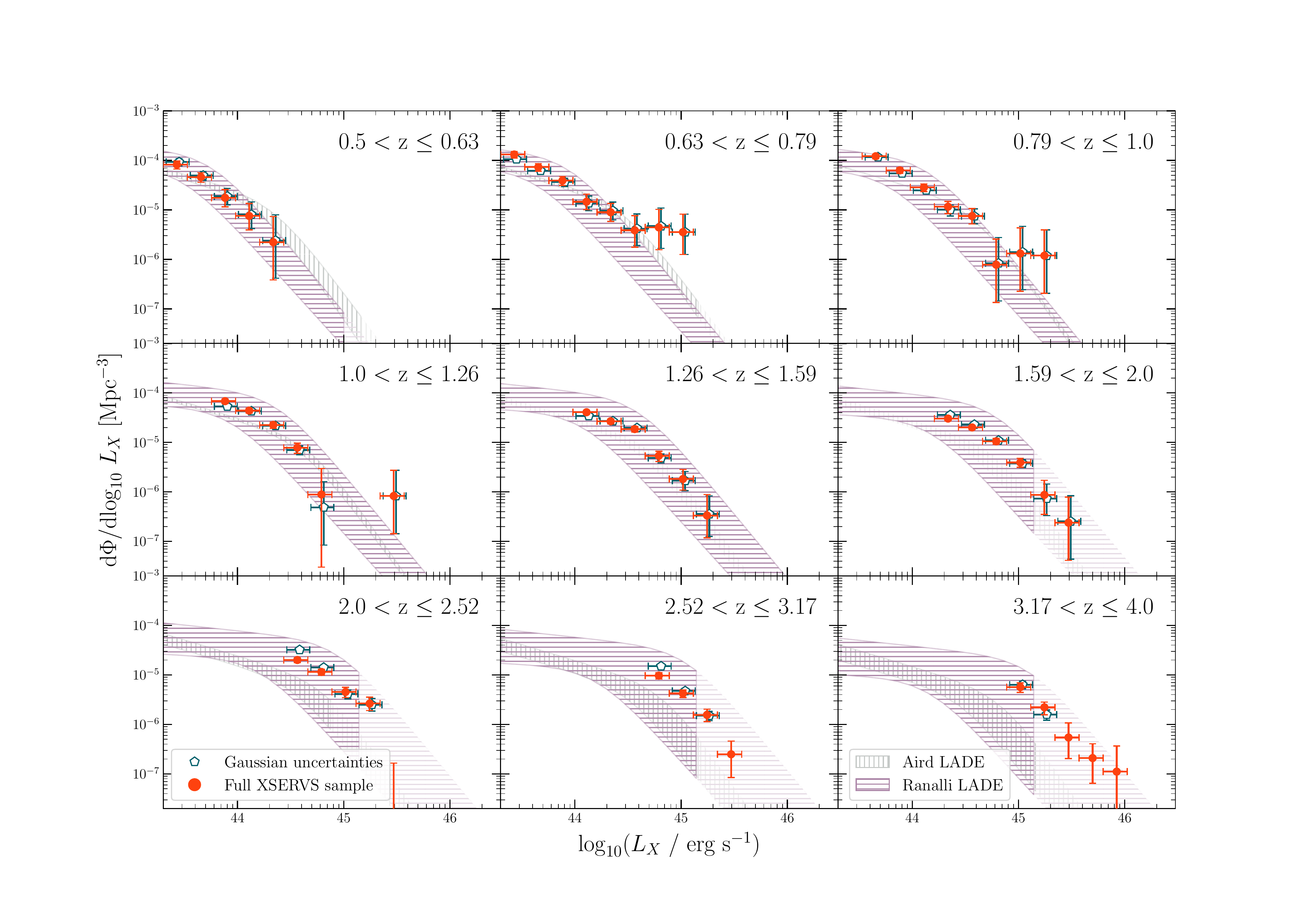}
\caption{XLF with Gaussian uncertainties on the photometric redshifts.}
\label{fig:lf_x_gauss}
\end{figure}

Figs.~\ref{fig:lf_ir_gauss} and \ref{fig:lf_x_gauss} show the IRLF and XLF, respectively, using Gaussian uncertainties for the photometric redshifts.
We find that the LF is relatively unchanged using these photometric redshift uncertainties.

\end{document}